
%
%
%
%
\ifx\epsffile\undefined\message{(FIGURES WILL BE IGNORED)}
\def\insertfig#1#2{}
\else\message{(FIGURES WILL BE INCLUDED)}
\def\insertfig#1#2{{{\baselineskip=4pt
\midinsert\hrule\centerline{\epsffile{#2}}{{#1}}\medskip\hrule\endinsert}}}
\fi

\input harvmac
%
%
%
%
\ifx\answ\bigans
\else
\output={
  \almostshipout{\leftline{\vbox{\pagebody\makefootline}}}\advancepageno
}
\fi
%
%
%

%
%

%
%
\def\UCSD#1#2{\noindent#1\hfill #2%
\bigskip\supereject\global\hsize=\hsbody%
\footline={\hss\tenrm\folio\hss}}
%
%
\def\abstract#1{\centerline{\bf Abstract}\nobreak\medskip\nobreak\par #1}
%
%
%
%
\edef\tfontsize{ scaled\magstep3}
 \tfontsize  \tfontsize
 \tfontsize \font\titlei=cmmi10 \tfontsize
\font\titleis=cmmi7 \tfontsize \font\titleiss=cmmi5 \tfontsize
\font\titlesy=cmsy10 \tfontsize \font\titlesys=cmsy7 \tfontsize
\font\titlesyss=cmsy5 \tfontsize  \tfontsize
\skewchar\titlei='177 \skewchar\titleis='177 \skewchar\titleiss='177
\skewchar\titlesy='60 \skewchar\titlesys='60 \skewchar\titlesyss='60
%
%
%
%
%
\def\inv{^{\raise.15ex\hbox{${\scriptscriptstyle -}$}\kern-.05em 1}}
\def\lbar{{\lower.35ex\hbox{$\mathchar'26$}\mkern-10mu\lambda}} 

%
%
%
%
\def\slash#1{\rlap{$#1$}/} 
\def\dsl{\,\raise.15ex\hbox{/}\mkern-13.5mu D} 
\def\delsl{\raise.15ex\hbox{/}\kern-.57em\partial}
\def\Ksl{\hbox{/\kern-.6000em\rm K}}
\def\Asl{\hbox{/\kern-.6500em \rm A}}
\def\Dsl{\hbox{/\kern-.6000em\rm D}} 
\def\Qsl{\hbox{/\kern-.6000em\rm Q}}
\def\gradsl{\hbox{/\kern-.6500em$\nabla$}}
%
%
\def\lspace{\ifx\answ\bigans{}\else\qquad\fi}
\def\lbspace{\ifx\answ\bigans{}\else\hskip-.2in\fi} 
%
%
\def\boxeqn#1{\vcenter{\vbox{\hrule\hbox{\vrule\kern3pt\vbox{\kern3pt
        \hbox{${\displaystyle #1}$}\kern3pt}\kern3pt\vrule}\hrule}}}
%
%
\def\mbox#1#2{\vcenter{\hrule \hbox{\vrule height#2in
\kern#1in \vrule} \hrule}}
%
%
%
%

   \def\CL{{\cal L}}
  \def\CO{{\cal O}}

%
%
%
%
%

%

\def\bar#1{\overline{#1}}
\def\vev#1{\left\langle #1 \right\rangle}
\def\bra#1{\left\langle #1\right|}
\def\ket#1{\left| #1\right\rangle}
\def\abs#1{\left| #1\right|}

\def\darr#1{\raise1.5ex\hbox{$\leftrightarrow$}\mkern-16.5mu #1}

%
%
\def\half{{\textstyle{1\over2}}} 
\def\frac#1#2{{\textstyle{#1\over #2}}} 
%
%
%
%

%
%
%
%

%
%
\def\ltap{\ \raise.3ex\hbox{$<$\kern-.75em\lower1ex\hbox{$\sim$}}\ }
\def\gtap{\ \raise.3ex\hbox{$>$\kern-.75em\lower1ex\hbox{$\sim$}}\ }
\def\gl{\ \raise.5ex\hbox{$>$}\kern-.8em\lower.5ex\hbox{$<$}\ }
\def\roughly#1{\raise.3ex\hbox{$#1$\kern-.75em\lower1ex\hbox{$\sim$}}}
%
%
\def\ie{\hbox{\it i.e.}}        
        
\def\etal{\hbox{\it et al.}}

\def\np#1#2#3{Nucl. Phys. B{#1} (#2) #3}
\def\pl#1#2#3{Phys. Lett. {#1}B (#2) #3}
\def\prl#1#2#3{Phys. Rev. Lett. {#1} (#2) #3}
\def\physrev#1#2#3{Phys. Rev. {#1} (#2) #3}

\relax

\def\wmunu{W^{\mu\nu}}
\def\tmunu{T^{\mu\nu}}
\def\rarr#1{\raise1.5ex\hbox{$\rightarrow$}\mkern-16.5mu #1}
\def\larr#1{\raise1.5ex\hbox{$\leftarrow$}\mkern-16.5mu #1}
\def\bfrac#1#2{{{#1}\over{#2}}}
\noblackbox
\def\lqcd{\Lambda_{\rm QCD}}

\centerline{{\titlefont{Inclusive Semileptonic $B$ and}}}
\bigskip
\centerline{{\titlefont{Polarized $\Lambda_b$ Decays from QCD}}}
\bigskip\medskip
\centerline{Aneesh V.~Manohar}
\smallskip
\centerline{{\sl Department of Physics, University of California at San
Diego, La Jolla, CA 92093}}
\bigskip
\centerline{Mark B.~Wise}
\smallskip
\centerline{{\sl California Institute of Technology, Pasadena, CA 91125}}
\vfill
\abstract{The differential decay spectrum $d\Gamma/dE_e dq^2$ for the
semileptonic decay of an unpolarized hadron containing a b-quark, and
the decay spectrum $d\Gamma/dE_e dq^2 d\!\cos\theta$ for polarized
$\Lambda_b$ decay are computed to second order in the $1/m_b$ expansion.
Most of the $1/m_b^2$ corrections have a simple physical interpretation,
which is discussed in detail. The implications of the results for the
determination of $V_{ub}$ are discussed. The decay spectra for
semileptonic decay of hadrons containing a c-quark are also given. There
is a subtlety in the use of the equations of motion at order $1/m_b^2$
in the heavy quark expansion which is explained.}
\vfill
\UCSD{\vbox{\hbox{UCSD/PTH 93-14}
\hbox{CALT-68-1883}\hbox{hep-ph/9308246}}}{August 1993}

\newsec{Introduction}
The inclusive lepton spectrum from semileptonic $B$ decays has undergone
intensive experimental and theoretical study.  An understanding of it
provides information on the weak mixing angles $V_{cb}$ and $V_{ub}$.
Recently there has been considerable progress in understanding the
theory of these decays.  Chay, Georgi and Grinstein~\ref\chay{J. Chay,
H. Georgi, and B. Grinstein, \pl{247}{1990}{399}}\ showed that
inclusive semileptonic $B \rightarrow X e\bar\nu_e$ decay can be treated
in a fashion similar to deep inelastic scattering.  Using a two step
process that consisted of first using an operator product expansion and
then a transition to the heavy quark effective theory, they showed that
$d\Gamma/dq^2 d E_e$ ($q^2 = (p_e + p_\nu)^2$)
when suitably averaged over $E_e$ is calculable.  Their leading order
result agrees with the free b-quark decay picture.  The full power of
their method becomes apparent when corrections to the leading
order result are discussed.  These corrections are of two types, perturbative
$\alpha_s (m_b)$ corrections and nonperturbative corrections suppressed
by powers of $m_b$.  Chay \etal\ pointed out that there are no
nonperturbative corrections of order $\lqcd/m_b$. In this paper, we
will compute the $\lqcd^2/m_b^2$ corrections. The order $\alpha_s(m_b)$
corrections have been computed previously~\ref\cabibbo{N. Cabibbo, G. Corb\`o
and L. Maiani, \np{155}{1979}{93}}\ref\ali{A. Ali and E. Pietarinen,
\np{154}{1979}{519}}\ref\corbo{G. Corb\`o,
\np{212}{1983}{99}}\ref\accmm{G.~Altarelli, N.~Cabibbo, G.~Corb\`o,
L.~Maiani, and G.~Martinelli, \np{208}{1982}{365}}. The results of this
paper can be combined with refs.~\cabibbo---\accmm\ to give the
inclusive semileptonic decay including all corrections to order
$\lqcd^2/m_b^2$ and $\alpha_s(m_b)$.

At leading order in $\alpha_s(m_b)$ and $\lqcd/m_b$ the
differential $B\rightarrow X_u e\bar\nu_e$ semileptonic decay rate is
\eqn\Ii{
{d\Gamma\over dq^2 dE_e} = {|V_{ub}|^2 G_F^2 m_b^2\over 8\pi^3}
\left[{2E_e\over m_b} - {q^2\over m_b^2}\right] \left[1 + {q^2\over m_b^2}
- {2E_e\over m_b}\right] \theta (2E_em_b - q^2).
}
In the approach of Chay \etal\ the b-quark mass appearing in eq.~\Ii\
has a precise meaning that is provided by the heavy quark effective
theory. This is necessary for the statement that there are no
$\lqcd/m_b$ corrections to have content. In the heavy quark effective
theory the strong interactions of a bottom quark with four velocity $v$
are given by the Lagrange density~\ref\geh{H. Georgi,
\pl{240}{1990}{447}\semi E. Eichten and B. Hill, \pl{234}{1990}{511}}
\eqn\Iii{
{\cal L} = \bar b_v\,( i v \cdot D)\, b_v + \ldots.
}
In eq.~\Iii\ the bottom quark field $b_v$ satisfies the constraint
$\slash v b_v = b_v$ and the ellipsis denote terms suppressed by
powers of $1/m_b$.  The relationship between the b-quark field in the
effective theory and the ``full QCD'' b-quark field is
\eqn\Iiii{
 b_v = {(1 + \slash v)\over 2} e^{im_b v\cdot x} b + \ldots,
}
where the ellipsis again denote terms suppressed by powers of $1/m_b$.
It is the same $m_b$ that appears in eq.~\Iiii\ that is used in eq.~\Ii.
With the heavy quark effective theory given by eq.~\Iii\ (\ie\
no mass term for $b_v$) $m_b$ is a physical quantity that can, at
least in principle, be determined experimentally.  For example, the form
factors for the exclusive decay $\Lambda_b \rightarrow  \Lambda_c e\bar
\nu_e$ depend on $\bar\Lambda = M_{\Lambda_{b}} - m_b$~\ref\ggw{H.
Georgi, B. Grinstein and M.B. Wise, \pl{252}{1990}{456}}\ and the
determination of $\bar \Lambda$ from a detailed study of this decay
together with the measured $\Lambda_b$ mass gives the b-quark mass to be
used in eq. (1.1).

Bigi, Shifman, Uraltsev and Vainshtein~\ref\bsuv{I.I.~Bigi, M.~Shifman,
N.G.~Uraltsev and A.I.~Vainshtein, \prl{71}{1993}{496}}\ have
performed an analysis of the
$\lqcd^2/m_b^2$ nonperturbative corrections to the lepton energy
spectrum $d\Gamma/dE_e$ in semileptonic B-meson decay.  Bigi \etal\
found that these corrections are determined by the two local matrix
elements $<B(v)|\bar b_v (iD)^2 b_v| B(v)>/2m_b^2$ and $<B(v)| g \bar
b_v \sigma^{\mu\nu} G_{\mu\nu} b_v |B(v)>/4m_b^2$.  The later is
fixed by the measured value of the $B^* - B$ mass difference.

In this paper we extend the results of Bigi \etal\ and Chay \etal, and compute
the $\lqcd^2/m_b^2$ corrections to the fully
differential decay distribution $d\Gamma/dq^2 dE_e$ for an unpolarized hadron
$H_b$ containing a b-quark.  We also consider
inclusive polarized semileptonic decay for the special case of the
$\Lambda_b$. $\Lambda_b$'s produced in $Z^0$ decays are expected to be
polarized~\ref\falkpeskin{A.F. Falk and M.E. Peskin, SLAC-PUB-6311
(1993)\semi T. Mannel and G.A. Schuler, \pl{279}{1992}{194}\semi
F.E. Close, J. Korner, R.J.N. Phillips, and D.J. Summers, J. Phys. G18
(1992) 1760}.
The differential decay distribution for a polarized  $\Lambda_b$ has the form
\eqn\Iiv{
 {d\Gamma\over dq^2 dE_e d\!\cos\theta} = A(E_e, q^2) + B(E_e, q^2)
\cos\theta,
}
where $\theta$ is the angle between the electron direction and the
$\Lambda_b$ spin in the rest frame of the $\Lambda_b$.  For reasons
similar to those given by Chay \etal\ in the case of $B$-meson decay,
there are no $\lqcd/m_b$ nonperturbative corrections to the
differential decay distribution in eq.~\Iiv.  The
$\lqcd^2/m_b^2$ corrections to $A(E_e, q^2)$ are similar
to those computed for $d\Gamma/dq^2 dE_e$ in $B$-meson decay.  One
simply replaces $B$-meson matrix elements by $\Lambda_b$ matrix elements
and sets the $\Lambda_b$ matrix element of
$\bar b_v g \sigma^{\mu\nu} G_{\mu\nu} b_v$ to zero.  However, we
find that the $\lqcd^2/m_b^2$ corrections to $B(E_e, q^2)$ are not
characterized by just $<\Lambda_b (v,s)|\bar b_v (iD)^2 b_v|\Lambda_b
(v,s)>/2m_b^2$.  The normalization of $B(E_e,q^2)$ involves another order
$\lqcd^2/m_b^2$ correction which arises because the heavy quark spin is
renormalized at order $1/m_b^2$.

We examine the physical interpretation of the $\lqcd^2/m_b^2$
corrections.  Most of the corrections (\ie\ all of those involving the
matrix element of $\bar b_v (iD)^2 b_v$ and some of those involving the
matrix element of $\bar b_v g \sigma^{\mu\nu} G_{\mu\nu} b_v$) can
be interpreted as arising from the fact that in the bound state the
b-quark has an effective mass that differs from $m_b$ and an effective
four velocity that differs from $v^\mu$.  These differences arise, for
example, from the motion of the b-quark in the hadron rest frame.
Corrections of this type are similar in spirit to those included in models for
inclusive semileptonic B-meson decay~\cabibbo---\accmm.

Section 2 contains a discussion of the kinematics relevant for
semileptonic $H_b$ and polarized $\Lambda_b$ decay.  The operator
product expansion and the transition to the heavy quark effective theory
are discussed in Section 3. This section contains a lengthy discussion of the
computation. Readers not interested in the details
are advised to skip this section entirely. Section 4 contains a brief
discussion on
the use of equations of motion in time ordered products.  It clears up
some confusion on this subject that occurred in the previous
literature.  Section 5 gives the differential decay rates for
unpolarized $H_b$
semileptonic decay and polarized $\Lambda_b$ decay. Section 6 is
concerned with the physical interpretation of the
$\lqcd^2/m_b^2$ corrections.
Section 7 gives differential decay
rates for unpolarized hadrons $H_c$ containing a c-quark and polarized
$\Lambda_c$ semileptonic decay.  Section
8 discusses the prediction for $d\Gamma(B \rightarrow X_u e\bar
\nu_e)/dq^2 dE_e$ near the boundary of the Dalitz plot.  This region is
important for the determination of $V_{ub}$.  Particular attention is
paid to the size of the region of $E_e$ that must be averaged over
before experimental results can be compared with theory.  Numerical
estimates and plots of the lepton spectrum are given in section~9, and
concluding remarks are given in Section 10.

\newsec{Kinematics}

The semileptonic decay of a b-quark is due to the weak hamiltonian
density
\eqn\IIi{
H_W = -  V_{jb}\ {4 G_F\over \sqrt 2}\ \bar q_j \gamma^\mu P_L b\ \bar e
\gamma_\mu P_L \nu_e = -  V_{jb}\ {4 G_F\over \sqrt 2}\ J^\mu_j J_{\ell \mu},
}
where $P_L$ is the left handed projection operator $\half(1-\gamma_5)$.
$J^\mu_j$ and $J^\mu_\ell$ are the hadronic and
leptonic currents, respectively. The final quark $q_j$ can be either a u- or a
c-quark. The inclusive differential decay rate for a hadron $H_b$ containing a
b-quark
to decay semileptonically, $H_b\rightarrow X_{u,c}\, e\bar \nu_e$ is determined
by the hadronic tensor
\eqn\IIiii{
\wmunu_j = \left(2\pi\right)^3\sum_X \delta^4\left(
p_{H_b}-q-p_X\right)\langle H_b (v,s)|J^{\mu\,\dagger}_j\ket{X}
\bra{X}J^\nu_j\ket{H_b(v,s)},
}
where $j=u,c$. The spin $J$ hadron state $\ket{H_b(v,s)}$ is normalized to
$v^0$, instead of to the usual relativistic normalization of $2 M_{H_b}v^0$ as
this is more convenient for the heavy quark expansion.
$W^{\mu\nu}$ can be expanded in
terms of five form factors if one spin-averages over the initial state,
\eqn\IIiv{
\wmunu=-g^{\mu\nu} W_1 + v^{\mu}v^{\nu} W_2 - i
\epsilon^{\mu\nu\alpha\beta} v_\alpha q_\beta W_3 + q^\mu q^\nu W_4 +
\left(q^\mu v^\nu + q^\nu v^\mu\right) W_5,
}
where $v$ is the velocity of the initial hadron, defined by
\eqn\IIv{p_{H_b} = M_{H_b}
v^\mu.
}
$W_1$ and $W_2$ have mass dimension $-1$, $W_3$ and $W_5$ have mass
dimension $-2$, and $W_4$ has mass dimension $-3$.
(The form factor $W_6$ of ref.~\chay\ vanishes by
time reversal invariance.)
The form factors are functions of the invariants $q^2$ and  $q\cdot
v$, and will also depend on the initial hadron $H_b$ and the final quark
mass $m_j$. The difference between the heavy quark mass $m_b$ and the
hadron mass
$M_{H_b}$ will be important in our analysis, so we have chosen to write the
form factors in terms of $q$ rather than the rescaled $\hat q = q/m_b$
used in Ref.~\chay.\foot{We will use $m$ to denote quark masses
and $M$ to denote hadron masses.} The spin averaged differential semileptonic
decay rate is
\eqn\IIvi{
{d\Gamma\over dq^2\, dE_e\, dE_\nu} = {\abs{V_{jb}}^2\, G_F^2\over 2
\pi^3}\left[ W_1 q^2 + W_2 \left(2 E_e E_\nu - \half q^2\right) + W_3
q^2\left(E_e - E_\nu\right)\right],
}
where $E_e$ and $E_\nu$ are the electron and neutrino energies in the
$H_b$ rest frame, $q^2$ is the
invariant mass of the lepton pair, and the kinematic variables are to be
integrated over the region $q^2 \le 4 E_e E_\nu$.  The terms proportional to
$q^\mu$ or $q^\nu$ in eq.~\IIiv\ do not contribute to the decay rate if one
neglects the electron mass.

The polarized $\Lambda_b$ has in addition to the five form-factors in
eq.~\IIiv, nine spin-dependent form factors which are
\eqn\IIvii{\eqalign{
W^{\mu\nu}_S &= - q\cdot s\Bigl[-g^{\mu\nu} G_1 + v^{\mu}v^{\nu} G_2 - i
\epsilon^{\mu\nu\alpha\beta} v_\alpha q_\beta G_3 + q^\mu q^\nu G_4 \cr
&\qquad +\left(q^\mu v^\nu + q^\nu v^\mu\right) G_5\Bigr] +
\left(s^\mu v^\nu + s^\nu v^\mu\right) G_6 +
\left(s^\mu q^\nu + s^\nu q^\mu\right) G_7 \cr
&\qquad
+i \epsilon^{\mu\nu\alpha\beta}v_\alpha s_\beta G_8
+i \epsilon^{\mu\nu\alpha\beta}q_\alpha s_\beta G_9.
}}
The identity
$$      g^{\mu\nu} \epsilon^{\alpha\beta\lambda\sigma}
-g^{\mu\alpha} \epsilon^{\nu\beta\lambda\sigma} + g^{\mu\beta}
\epsilon^{\alpha\nu\lambda\sigma} -g^{\mu\lambda}
\epsilon^{\nu\alpha\beta\sigma} +g^{\mu\sigma}
\epsilon^{\nu\alpha\beta\lambda}= 0, $$
has been used to eliminate terms in $W^{\mu\nu}_S$ of the form
$i(q^\mu \epsilon^{\nu\alpha\beta\lambda} v_\alpha q_\beta s_\lambda
-  (\mu \rightarrow \nu))$ and $ i(v^\mu \epsilon^{\nu\alpha\beta\lambda}
v_\alpha q_\beta s_\lambda - (\mu \rightarrow \nu)).$
The form factors $G_4$, $G_5$, $G_7$ do not contribute to the decay rate if the
lepton mass is neglected. The differential decay rate is
\eqn\IIci{\eqalign{
&{d\Gamma \over dq^2 dE_e dE_\nu d\!\cos\theta} = ... + {|V_{jb} |^2 G_F^2\over
4\pi^3} \cos \theta\ \times\cr
&\Bigg[\left(G_1 q^2 + G_2 \left[2E_e E_\nu - {1\over 2} q^2\right] + G_3 q^2
\left[E_e -  E_\nu\right]\right)\left(E_e + E_\nu - q^2/2 E_e
\right)\cr
&\qquad + G_6 \left(q^2 - 4E_e E_\nu\right)
- G_8 q^2 - G_9 q^2 \left(E_e - E_\nu + q^2/2E_e\right)\Bigg],
}}
where $\theta$ is the angle between the electron three momentum and the
$\Lambda_b$ spin vector in the $\Lambda_b$'s rest frame.  The ellipsis
in eq.~\IIci\ denotes the part independent of $\cos\theta$ and is
one half the expression in eq.~\IIvi\ so that integration over
$\cos\theta$ reproduces the unpolarized decay rate.

The form factors in $\wmunu$ and $\wmunu_S$ are given by the discontinuities
across a cut of the amplitudes $\tmunu$ and $\tmunu_S$,
\eqn\IIviii{\eqalign{
\tmunu&=-i\int d^4 x\ e^{-i q \cdot x}{1\over 2J+1} \sum_s
\langle H_b (v,s)| T\left( J^{\mu\,\dagger}\left(x\right)
J^\nu\left(0\right)\right) \ket{H_b(v,s)}\cr
&=-g^{\mu\nu} T_1 + v^{\mu}v^{\nu} T_2 - i
\epsilon^{\mu\nu\alpha\beta} v_\alpha q_\beta T_3 + q^\mu q^\nu T_4 +
\left(q^\mu v^\nu + q^\nu v^\mu\right) T_5.
}}
It is easy to see that ${\rm Im}\, T^{\mu\nu} =-\pi W^{\mu\nu}$ by
inserting a complete set of states between the currents.  $T_S^{\mu\nu}$ for
the polarized $\Lambda_b$ is defined similarly and has nine additional
spin-dependent form-factors,
\eqn\IIix{\eqalign{
T^{\mu\nu}_S &= - q\cdot s\Bigl[-g^{\mu\nu} S_1 + v^{\mu}v^{\nu} S_2 - i
\epsilon^{\mu\nu\alpha\beta} v_\alpha q_\beta S_3 + q^\mu q^\nu S_4 \cr
&\qquad +\left(q^\mu v^\nu + q^\nu v^\mu\right) S_5\Bigr] +
\left(s^\mu v^\nu + s^\nu v^\mu\right) S_6 +
\left(s^\mu q^\nu + s^\nu q^\mu\right) S_7 \cr
&\qquad
+i \epsilon^{\mu\nu\alpha\beta}v_\alpha s_\beta S_8
+i \epsilon^{\mu\nu\alpha\beta}q_\alpha s_\beta S_9.
}}

\insertfig{\centerline{Figure 1}\medskip
The analytic structure of $T^{\mu\nu}$ in the complex $q\cdot
v$ plane for fixed timelike $q^2$. The contour $C$ encircles the cut
relevant for semileptonic b-decay. The other cuts extend to infinity,
and correspond to other physical processes. Here $q\equiv\sqrt{q^2}$.}
{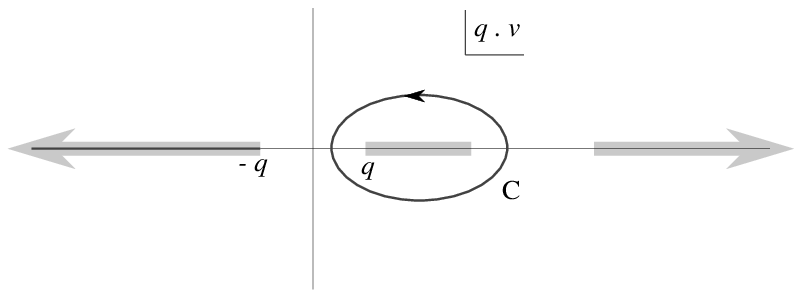}

The analytic structure
of $\tmunu$ (or $T_S^{\mu\nu}$) as a function of $q\cdot v$ for fixed $q^2$
was given in ref.~\chay\ and is show in fig~1.
There is a cut  from $\sqrt {q^2} \le q\cdot v \le \left({1\over
2M_{H_b}}\right)\left(M_{H_b}^2 +  q^2 - M_j^2\right)$, where $M_j$ is the mass
of the lightest hadron  containing the final state quark $q_j$.
The discontinuity across this cut  (which will be called the physical cut)
gives $\wmunu$ for semileptonic $H_b$  decay. In addition there are cuts
along the real axis for $M_{H_b} \, q \cdot v \ge   \half\left((2 M_{H_b} +
M_j)^2-q^2-M_{H_b}^2\right)$ corresponding to the physical process
$e\bar\nu_e H_b \rightarrow X$ (where $X$ contains two b-quarks),
and for $v\cdot q \le -\sqrt {q^2}$ corresponding to the physical
process $e^+\nu_e H_b\rightarrow X$. There is
no cut for $-\sqrt {q^2} \le q \cdot v \le \sqrt
{q^2}$, since $ (q^0)^2 > q^2$ for any physical state,
so that, in general, the physical cut is separated from the other two
cuts along the real axis. (This disagrees with ref.~\chay.)  For $M_j = 0$
and $q^2 = M_{H_b}^2$ the two cuts on the positive real axis are not
separated.  The end of the physical cut at $v\cdot q = M_{H_b}$ coincides with
the beginning of the second cut.  Similarly for $q^2 = 0$ the cut on the
negative real axis ends at the same point the physical cut begins.

The amplitude $\tmunu$ can be computed in QCD perturbation theory away
from the cuts along the real axis. Provided $q^2$ is not too near zero
or $M_{H_b}^2$ (for $q_j = u$), the value of the amplitude in the
physical region can be obtained by performing a contour integral
along the closed contour $C$ shown in fig.~1 that stays away from the
cuts in the complex $q\cdot v$ plane.

\newsec{The Operator Product Expansion}

The amplitude $\tmunu$ can be computed reliably in perturbative QCD in a
region that is far from the cuts and therefore free of infrared
singularities. The time ordered product
\eqn\IIIi{
-i\int d^4 e^{-i q \cdot x} T(J^{\mu\dagger} J^\nu)
}
can be computed using an operator product expansion in terms of
operators involving b-quark fields in the heavy quark effective field
theory. The coefficients of the operators in the operator
product expansion are determined by evaluating the matrix element of the
time-ordered product between quark and gluon states. Once the operator product
expansion has been computed, one can compute $\tmunu$ by taking the
matrix element of the operator product expansion between hadron
states. We will compute the matrix element between unpolarized hadron states
to determine the form-factors $T_i$, and we will take the matrix element
between polarized $\Lambda_b$ states to determine the spin-dependent
form-factors $S_i$.  The operator product expansion can be written as an
expansion in inverse powers of $m_b$. The expansion of $\tmunu$ to order
$1/m_b^2$ can be written in terms of gauge invariant operators of dimension
less than or equal to five. The operators will involve the $b$ field,
covariant derivatives $D$, and the gluon field strength tensor $G^{\mu\nu}$.
The coefficients of the operators involving $b$ and $D$ are determined
by taking quark matrix elements of eq.~\IIIi, and the coefficient of
operators involving $G^{\mu\nu}$ are determined by taking gluon matrix
elements of eq.~\IIIi. We only calculate the form-factors $T_{1-3}$,
$S_{1-3}$, $S_6$, $S_8$ and $S_9$ which are the only ones that contribute to
semileptonic b-decay when the mass of the lepton in the final state is
neglected.

\insertfig{\centerline{Figure 2}\medskip
The leading term in the operator product expansion. Fig.~2(b) does not
contribute to semileptonic b-decay.}{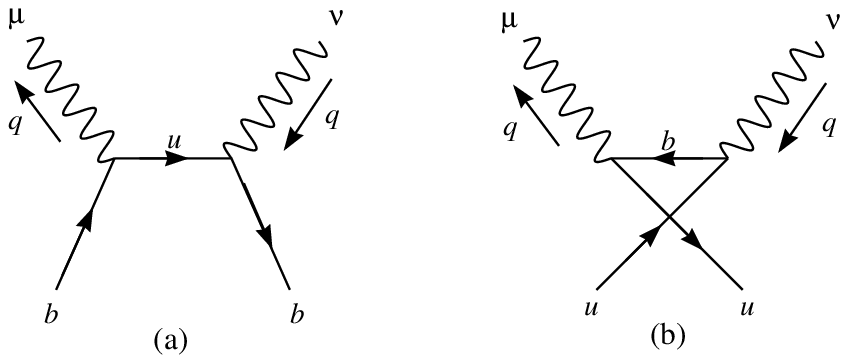}

The quark matrix element of eq.~\IIIi\ between b-quark states with
momentum $m_b v + k$ is
\eqn\IIIii{
{1\over (m_b v - q  + k )^2 - m_j^2 + i \epsilon}\
\bar u\ \gamma^\mu\, P_L\ \left(m_b\slash v - \slash q +
\slash k + m_j\right)
\ \gamma^\nu\, P_L\ u ,
}
from fig.~2(a), where $u$ is the quark spinor.
The crossed diagram of fig.~2(b)  has no singularities inside
the contour $C$ of fig.~1, and does not contribute to semileptonic
b-decay. The matrix element eq.~\IIIii\ can be simplified using the identity
$$
\gamma^\mu \gamma^\alpha \gamma^\nu = g^{\mu\alpha} \gamma^\nu +
g^{\nu\alpha} \gamma^\mu - g^{\mu\nu} \gamma^\alpha + i
\epsilon^{\mu\nu\alpha\beta}\gamma_\beta \gamma_5.
$$
The $1/m_b$ expansion is given by expanding eq.~\IIIii\ in a power series.
The momentum $q$ can be of order $m_b$, but $k$ is only of order $\lqcd$. Thus
each factor of $k$ in the expansion of eq.~\IIIii\ corresponds to a $1/m_b$
suppression. The factors of $k$ in the matrix element will become factors of
$iD$ in the operator product expansion, so each factor of $iD$ corresponds to a
$1/m_b$ suppression.

\subsec{The $k^0$ Terms}

The order $k^0$ term in the expansion of eq.~\IIIii\ is
\eqn\IIIiii{\eqalign{
{1\over\Delta_0} &\bar u\, \Bigl\{\  \left(m_b v - q\right)^\mu
\gamma^\nu + \left(m_b v - q\right)^\nu \gamma^\mu -  \left(m_b \slash v
- \slash q\right) g^{\mu\nu}\cr
&\qquad - i \epsilon^{\mu\nu\alpha\beta} \left(m_b v - q\right)_\alpha
\gamma_\beta\ \Bigr\} P_L\, u,
}}
where
\eqn\IIIiv{
\Delta_0 = (m_b v - q  )^2 - m_j^2 + i\epsilon.
}
The matrix elements of the operators $\bar b\, \gamma^\lambda \,b$ and $\bar
b\, \gamma^\lambda \gamma_5\, b$ between b-quark states are $\bar u\,
\gamma^\lambda\, u$
and $\bar u\, \gamma^\lambda \gamma_5 \,u$ respectively, so the operator
product
expansion is obtained by replacing $\bar u$ and $u$ in eq.~\IIIiii\ by
the fields $\bar b$ and $b$ respectively,
\eqn\IIIv{\eqalign{
{1\over\Delta_0} &\Bigl\{ \left(m_b v - q\right)^\mu g^{\nu\lambda}
+ \left(m_b
v - q\right)^\nu g^{\mu\lambda} -
\left(m_b  v -  q\right)^\lambda
g^{\mu\nu}\cr &\qquad - i \epsilon^{\mu\nu\alpha\lambda} \left(m_b v
- q\right)_\alpha\Bigr\}\ \bar b\, \gamma_\lambda P_L\, b.
}}
The spin averaged matrix element $\sum_s \bra{H_b (v,s)} \bar b\,
\gamma^\lambda \gamma_5 \, b\ket{H_b (v,s) }$ between
hadrons $H_b$ of velocity $v$ is zero. The matrix element
$$
\bra{H_b (v,s) }\bar b\, \gamma^\lambda \, b\ket{H_b (v,s)}=v^\lambda,
$$
to all orders in $1/m_b$, since b-quark number
is an exact symmetry of QCD. (This relation is true because we
have normalized our hadron states to $v^0$, and defined $v$ in eq.~\IIv\
to be the velocity of the hadron $H_b$.)
Taking the matrix element of the operator product expansion between
spin-averaged $H_b$ states, and comparing with eq.~\IIviii\ gives the
contribution to $\tmunu$ of $k^0$ terms in the expansion of eq.~\IIIii,
\eqn\IIIvi{\eqalign{ T_1^{(0)} &= {1\over 2\Delta_0}\left(m_b-q\cdot v \right),
\cr
\noalign{\smallskip}
T_2^{(0)} &= {1\over\Delta_0}\ m_b,\cr
\noalign{\smallskip}
T_3^{(0)} &= {1\over 2\Delta_0}.
}}
The $k^0$ terms in the operator product expansion give b-quark operators that
have zero derivatives. In principle, these operators could produce
contributions to $\tmunu$ of higher order in $1/m_b$ because of $1/m_b$
corrections to the matrix element of the operator between hadron states.
However, the only matrix element we need is the zero momentum transfer
matrix element of a conserved current, which has no $1/m_b$ corrections.

The matrix element of the operator $\bar b \gamma^\lambda \gamma_5 b$ between
polarized $\Lambda_b$ states does not vanish. At leading order (in
$1/m_b$) it is equal to the spin vector $s^\lambda$.
However, the axial current is a
generator of heavy quark symmetry~\ref\isgurwise{N. Isgur and M.B. Wise,
\pl{232}{1989}{113}}\ that is broken by $1/m_b$ terms in
the effective
Lagrangian. Thus the matrix element of the axial current between
polarized $\Lambda_b$ states is not equal to the spin four vector. The first
correction to the matrix
element is of second order in symmetry breaking~\ref\luke{M.E. Luke,
\pl{252}{1990}{447}}, so we write
\eqn\IIIvii{\eqalign{
\bra{\Lambda_b (v,s) } \bar b \gamma^\lambda \gamma_5 b
\ket{\Lambda_b (v,s) }
&= \left(1+\epsilon_b\right)\ \bar u(v,s) \gamma^\lambda \gamma_5 u(v,s),\cr
&= (1 + \epsilon_b) s^\lambda,
}}
where $\epsilon_b$ is a parameter which is of order $\lqcd^2/m_b^2$, and is
defined by eq.~\IIIvii. Substituting this into eq.~\IIIv\ gives the
spin-dependent form-factors from the $k^0$ terms
\eqn\IIIviii{
\eqalign{
S^{(0)}_1 &= -\bfrac12 (1+\epsilon_b) {1\over\Delta_0},\cr
\noalign{\smallskip}
S^{(0)}_6 &= -\bfrac12 (1+\epsilon_b) {m_b\over\Delta_0},
}\qquad
\eqalign{
S^{(0)}_8 &= \bfrac12 (1+\epsilon_b) {m_b\over\Delta_0},\cr
\noalign{\smallskip}
S^{(0)}_9 &= -\bfrac12 (1+\epsilon_b) {1\over\Delta_0},
}
}
$$      S_2^{(0)} = S_3^{(0)} = 0.$$

\subsec{The Order $k^1$ Terms: Spin Averaged Case}

The linear terms in $k$ in eq.~\IIIii,
\eqn\IIIix{\eqalign{
&{1\over \Delta_0}\bar u\, \Bigl\{k^\mu \gamma^\nu + k^\nu \gamma^\mu
-g^{\mu\nu}\slash k - i \epsilon^{\mu\nu\alpha\beta} k_\alpha\gamma_\beta
\Bigr\}\, P_L\, u\cr
\noalign{\smallskip}
&-{2k\cdot(m_b v - q)
\over \Delta_0^2} \bar u\, \Bigl\{ \left(m_b v - q\right)^\mu
\gamma^\nu + \left(m_b v - q\right)^\mu \gamma^\nu -  \left(m_b \slash v
- \slash q\right) g^{\mu\nu}\cr
&\qquad\qquad\qquad\qquad\qquad - i \epsilon^{\mu\nu\alpha\beta} \left(m_b v -
q\right)_\alpha \gamma_\beta\Bigr\} P_L\, u,
}}
produce operators in the operator product expansion with one derivative,
of the form $\bar b\, \gamma^\lambda \, i D^\tau\,
b$ and
$\bar b\, \gamma^\lambda \gamma_5\, i D^\tau\, b$. The matrix elements
of these operators need to be computed to first
order in $1/m_b$, since they contribute to terms that are already
suppressed by $1/m_b$. Unlike the current operator $\bar
b\,\gamma^\lambda\, b$ in
the $k^0$ terms, these operators will have $1/m_b$ corrections to their matrix
elements, and so can contribute terms to $\tmunu$ at order $1/m_b$ and
$1/m_b^2$.
The matrix element of $\bar b\, \gamma^\lambda \gamma_5\, i D^\tau\,
b$ vanishes between spin averaged $H_b$ states. Its contribution to the
polarized
$\Lambda_b$ decay amplitude is discussed in the next subsection.
The matrix element of $\bar b\, \gamma^\lambda \, i D^\tau\, b$ can be computed
in a $1/m_b$ expansion using the heavy quark effective theory.

The b-quark is represented by the velocity dependent b-quark field
$b_v$ in the heavy quark effective theory. Since the terms linear in $k$
are already of order $1/m_b$, we only need the relation between $b(x)$
and $b_v(x)$ to first order in $1/m_b$~\ref\fgleh{G.P. Lepage and B.A.
Thacker, Nucl. Phys. B4 (Proc. Suppl.) (1988) 199\semi
E. Eichten and B. Hill, \pl{243}{1990}{427}\semi
A.F.~Falk, B.~Grinstein, and M.E.~Luke, \np{357}{1991}{185}},
\eqn\IIIx{
b(x) = e^{-i m_b v\cdot x} \left[1 + {{i\, \Dsl}\over 2 m_b} \right] b_v(x).
}
The QCD lagrangian for the b-quark in the heavy quark effective theory
is
\eqn\IIIxi{
\CL = \bar b_v\, i v \cdot D\, b_v + \bar b_v\, {(i D)^2\over 2m_b}\,
b_v
- Z_b\,
\bar b_v\, { g G_{\alpha\beta}\sigma^{\alpha\beta}\over 4m_b}\, b_v +
\CO\left({1\over m_b^2}\right).
}
where $Z_b$ is a renormalization factor, with $Z_b(\mu=m_b)=1$. The operator
$\bar b_v\, (i D)^2\, b_v/2m_b $ is not renormalized because of
reparameterization invariance~\ref\reparaminv{M.E.~Luke and
A.V.~Manohar, \pl{286}{1992}{348}}.

Eq.~\IIIx\ gives the expansion of the operator in the effective theory
\eqn\IIIxii{\eqalign{
\bar b\, \gamma^\lambda\, i  D^\tau b &= \bar b_v\, \gamma^\lambda \,i
D^\tau b_v +
\bar b_v\, {{- i\, \larr \Dsl}\over 2 m_b} \gamma^\lambda\, i D^\tau\, b_v
+ \bar b_v\, \gamma^\lambda \, i  D^\tau  {{i\, \Dsl}\over 2 m_b}\, b_v ,\cr
&=v^\lambda \bar b_v\,i
D^\tau b_v +{1\over m_b}
\bar b_v\, iD^{(\lambda} \, i D^{\tau)}\, b_v
-{1\over 2m_b}\bar  b_v\, gG^{\alpha\tau}\sigma_\alpha{}^\lambda\, b_v ,\cr
}}
where we have used the relation $\bar b_v\, \gamma^\lambda b_v =
v^\lambda \bar b_v\, b_v$ (which follows from the constraint $\slash v
b_v=b_v$) and the commutator $[D^\mu, D^\nu] = ig G^{\mu\nu}$.  In eq.
(3.12) the brackets around indices denote that they are symmetrized.

The spin averaged matrix element of the first term of eq.~\IIIxii\ must have
the form
\eqn\IIIxiii{
{1\over (2J+1)} \sum_s \bra{H_b (v,s) } \bar b_v \,i  D^\tau\, b_v
\ket{H_b (v,s)} = A v^\tau ,
}
where $A$ is a constant to be determined.
Contracting both sides of
eq.~\IIIxiii\ with $v^\tau$ gives
\eqn\IIIiv{
A = \bra{H_b (v,s) } \bar b_v\,  i D \cdot v\, b_v
\ket{H_b (v,s) }.
}
The coefficient $A$ is zero at lowest order in $1/m_b$, since $(v\cdot
D)\,b_v=0$ is
the lowest order equation of motion in the heavy quark effective theory. Thus
the order  $k$ terms make no contribution to $\tmunu$ at
order $1/m_b$. The order $k^0$ terms also did not contribute to $\tmunu$ at
order $1/m_b$,  so there is no $1/m_b$ correction to $\tmunu$. This
reproduces the result of ref.~\chay.
The parameter $A$ is non-zero at first order in $1/m_b$,
\eqn\IIIxv{
A =-  \bra{H_b (v,s) }\left[
\bar b_v\, {(i D)^2\over 2m_b}\, b_v - Z_b\,
\bar b_v\, { g G_{\alpha\beta}\sigma^{\alpha\beta}\over 4m_b}\, b_v \right]
\ket{H_b (v,s) } ,
}
using the equations of motion from the Lagrangian eq.~\IIIxi\ to order
$1/m_b$. This $1/m_b$ value of $A$ contributes to $\tmunu$ at order $1/m_b^2$.

It is useful to define the dimensionless parameters
\eqn\paramdef{\eqalign{
E_b &\equiv -  \bra{H_b(v,s)}\left[ \bar b_v\, {(i
D)^2\over 2m_b^2}\, b_v - Z_b\,
\bar b_v\, { g G_{\alpha\beta}\sigma^{\alpha\beta}\over 4m_b^2}\, b_v \right]
\ket{H_b(v,s)} ,\cr
K_b &\equiv  -  \bra{H_b(v,s)} \bar b_v {(i
D)^2\over 2m_b^2}\, b_v\ket{H_b(v,s)},\cr
G_b &\equiv  Z_b\,\bra{H_b(v,s)}
\bar b_v\, { g G_{\alpha\beta}\sigma^{\alpha\beta}\over 4m_b^2}\, b_v
\ket{H_b(v,s)},
}}
to characterize the $1/m_b^2$ corrections, with $E_b=K_b+G_b$. $G_b$,
$K_b$ and $E_b$ can be thought of as the average value of the
spin-energy, the kinetic energy, and the total energy of the b-quark in
the hadron $H_b$, in units of $m_b$.
All three parameters are expected to be order
$\lqcd^2/m_b^2$. The operators in eq.~\paramdef\ are renormalization point
independent, since the $\mu$ dependence of $Z_b$ cancels the $\mu$
dependence of
$\bar b_v\, g G_{\alpha\beta}\sigma^{\alpha\beta}\, b_v$. The anomalous
dimensions of the operators only affect the relations between the
parameters for
different heavy quarks. For example, the parameters for $b$ and $c$ quarks are
related by
\eqn\IIIxvii{\eqalign{
m_c^2 K_c &= m_b^2 K_b,\cr
m_c^2 G_c/Z_c &= m_b^2 G_b/Z_b. }}
Since $Z_c(m_c)=1$ and $Z_b(m_b)=1$, the ratio $Z_b/Z_c$ is
given by the scaling of $\bar b_v\, g
G_{\alpha\beta}\sigma^{\alpha\beta}\, b_v$
between the scales $m_c$ and $m_b$, $Z_b/Z_c =
\left[\alpha_s(m_b)/\alpha_s(m_c)\right]^{9/25}$~\fgleh. The
gluon operator $G^{\mu\nu}$ in the operator product expansion occurs
through the
equations of motion, and through the commutator $\left[D^\mu,D^\nu\right]=ig
G^{\mu\nu}$. We will use $E_b$ to parameterize the matrix elements of the gluon
operators obtained using the equations of motion, and $G_b$ to
parameterize the matrix elements of the gluon operators obtained from
$\left[D^\mu,D^\nu\right]$. This distinction will be useful in sec.~6.
With this convention,
\eqn\IIIxviii{
A = m_b E_b,
}
and the matrix element of the first term of eq.~\IIIxiii\ can now be written as
\eqn\IIIxix{
{1\over (2J+1)} \sum_s \bra{H_b (v,s) } v^\lambda\bar b_v\,   i  D^\tau\, b_v
\ket{H_b (v,s)}= m_b E_b v^\lambda v^\tau.
}

The matrix element of the second term in eq.~\IIIxii\ must have the form
\eqn\IIIxx{
{1\over (2J+1)} \sum_s \bra{H_b (v,s)} \bar b_v\, iD^{(\lambda} \, i
D^{\tau)}\, b_v\ket{H_b (v,s) } = \left(B_1 g^{\lambda\tau}+B_2 v^\lambda
v^\tau\right).
}
This term has two covariant derivatives, so we only need its matrix
element to lowest order in $1/m_b$. The lowest order equation of motion
$(v\cdot D)\,b_v=0$ implies that $B_1+B_2=0$. Taking the trace and
comparing with
eq.~\paramdef\ gives
\eqn\IIIxxi{
{1\over 2J+1} \sum_s \bra{H_b(v,s)} \bar b_v\, iD^{(\lambda} \, i D^{\tau)}
\, b_v\ket{H_b(v,s)} = -{2 m_b^2 K_b\over 3}\left(g^{\lambda\tau}-
v^\lambda v^\tau\right).
}

The operator $\bar b_v\, g G_{\alpha\tau}\sigma^{\alpha\lambda}\, b_v$
in eq.~\IIIxii\ is renormalized at a scale $\mu=m_b$, since that is the scale
at
which the operator product expansion has been performed. We can therefore
multiply the operator by the renormalization factor $Z_b$, since $Z_b(m_b)=1$.
This makes the operator renormalization group invariant, and includes the
QCD scaling of the operator between $m_b$ and $\mu$. The matrix element of the
third term in eq.~\IIIxii\ must have the form
\eqn\IIIxxii{
{1\over 2J+1} \sum_s Z_b \bra{H_b(v,s)}
\bar b_v\, gG^{\alpha\tau}\sigma_\alpha{}^\lambda\, b_v
\ket{H_b(v,s)} =  \left(C_1 g^{\lambda\tau}+C_2 v^\lambda v^\tau\right).
}
Contracting both sides with
$v^\tau$, and using $\bar b_v\, \sigma^{\alpha\tau}v_\tau \, b_v=0$ gives
$C_1+C_2=0$. The trace gives
\eqn\IIIxxiii{{1\over 2J+1} \sum_s Z_b \bra{H_b(v,s)} \bar b_v\,
gG^{\alpha\tau}\sigma_\alpha{}^\lambda\, b_v \ket{H_b(v,s)} = {4
m_b^2 G_b\over
3}\left(g^{\lambda\tau} - v^\lambda v^\tau\right), }
on comparing with eq.~\paramdef.

Substituting eq.~\IIIxix, \IIIxxi\ and \IIIxxiii\ into
eq.~\IIIxii, and substituting the result into eq.~\IIIix\ gives
\eqn\IIIxxiv{\eqalign{
T_1^{(1)}&= m_b E_b\left[ {1\over2 \Delta_0} - {(m_b-q\cdot
v)^2\over\Delta_0^2}
\right] + {2 m_b\over 3} \left(K_b+G_b\right)\left[
-{1\over 2 \Delta_0} + { q^2 - (q\cdot v)^2 \over \Delta_0^2}\right],\cr
\noalign{\smallskip}
T_2^{(1)}&= m_b E_b\left[ {1\over\Delta_0} - 2 {m_b (m_b-q\cdot
v)\over\Delta_0^2}
\right]+{2 m_b\over 3} \left(K_b+G_b\right)\left[
{1\over \Delta_0} + {2 m_b q\cdot v \over \Delta_0^2}\right],\cr
\noalign{\smallskip}
T_3^{(1)}&= - m_b E_b\left[ {(m_b-q\cdot
v)\over\Delta_0^2}\right]-{2 m_b\over 3} \left(K_b+G_b\right)\left[
{ m_b-q\cdot v \over \Delta_0^2}\right].
}}

\subsec{The Order $k^1$ Terms: Polarized $\Lambda_b$ Case}

The spin-dependent form-factors arising from the order $k$ terms in the
operator product expansion of eq.~\IIIix\ are obtained by taking the matrix
element of the operator $\bar b\, \gamma^\lambda \gamma_5\, iD^\tau \,b$
between
polarized $\Lambda_b$ states. This operator did not contribute to the
spin-averaged matrix element between unpolarized $H_b$ states discussed in
the previous section. The method used to evaluate the matrix element is similar
to that used for the operator  $\bar b \,\gamma^\lambda \,iD^\tau\, b$  in the
previous subsection. The operator can be written in terms of the field $b_v$ of
the effective theory,
\eqn\IIIxxv{ \bar b \,\gamma^\lambda\gamma_5\, iD^\tau\, b = \bar b_v
\,\gamma^\lambda\gamma_5\, iD^\tau\, b_v + {1\over 2m_b} \bar b_v\, i\Dsl
\,\gamma^\lambda \gamma_5\,iD^\tau\, b_v +
{1\over 2m_b} \bar b_v\, \gamma^\lambda \gamma_5\,iD^\tau\, i\Dsl\,  b_v.
}
The matrix element of the first term of eq.~\IIIxxv\ between polarized
$\Lambda_b$ states has to have the form
\eqn\IIIxxvi{
\bra{\Lambda_b(v,s)}\bar b_v\, \gamma^\lambda\gamma_5\, iD^\tau\, b_v
\ket{\Lambda_b(v,s)} = A\ \bar u\, \gamma^\lambda \gamma_5\, u\ v^\tau,
}
by heavy quark spin-symmetry. Contracting both sides with $v^\tau$ and using
the equations of motion determines $A=m_b E_b$, where $E_b$ is defined in
eq.~\paramdef. The matrix element of the second and third terms in
eq.~\IIIxxv\  can be simplified by neglecting terms proportional to
$G^{\alpha\beta}$. The operator $G^{\alpha\beta}$ vanishes in any matrix
element
between $\Lambda_b$ states at zero recoil, since the light degrees
of freedom in the $\Lambda_b$ have spin-zero.  (That is why spin
symmetry can be used in eq.~\IIIxxvi\ even though we are including effects
of order $1/m_b$.)  The only vector that can be
constructed using the light degrees of freedom is $v^\mu$, and it is not
possible to construct a tensor that is antisymmetric in two indices from a
single vector.

The $1/m_b$ terms in eq.~\IIIxxv\ can be simplified using
$\gamma$-matrix algebra and neglecting $G^{\alpha\beta}$ to give
\eqn\IIIxxvii{
{1\over m_b} \bar b_v\, iD^{(\alpha} iD^{\tau)}\, (v_\alpha\gamma^\lambda-
v^\lambda\gamma_\alpha)\gamma_5 \, b_v ,
}
which is equal to
\eqn\IIIxxviii{
- {1\over m_b} \bar b_v\, iD^{(\alpha} iD^{\tau)}\,
v^\lambda\gamma_\alpha\gamma_5 \, b_v, }
using the equations of motion. The matrix element of eq.~\IIIxxviii\  between
polarized $\Lambda_b$ states is
\eqn\IIIxxix{
- {1\over m_b}\,\bra{\Lambda_b(v,s)}
 \bar b_v iD^{(\alpha} iD^{\tau)} v^\lambda\gamma_\alpha\gamma_5
b_v \ket{\Lambda_b(v,s)} = \bfrac 2 3 m_b K_b\, v^\lambda\, \bar u
\,\gamma^\tau\gamma_5\, u,
}
using heavy quark spin-symmetry, and the matrix element eq.~\IIIxxi.
Eqs.~\IIIxxvi, \IIIxxvii, and \IIIxxix\ imply that one can make the
substitution
\eqn\IIIxxx{ \bar b\, \gamma^\lambda \gamma_5\, iD^\tau\, b
\rightarrow m_b E_b\, s^\lambda v^\tau
+ \bfrac 2 3 m_b K_b \, v^\lambda s^\tau,
}
in eq.~\IIIix\ to obtain the spin-dependent form factors
\eqn\IIIxxxi{\eqalign{
S_1^{(1)} & = {1\over\Delta_0^2}\left(m_b-q\cdot v\right) \left(m_b E_b +
\bfrac23m_bK_b\right),\cr
\noalign{\smallskip}
S_2^{(1)} & = {4\over3\Delta_0^2} m_b^2 K_b,\cr
\noalign{\smallskip}
S_3^{(1)} & = {2\over 3\Delta_0^2}m_bK_b,\cr
\noalign{\smallskip}
S_6^{(1)} & = -{1\over2\Delta_0} \left(m_b E_b +
\bfrac23m_bK_b\right) +
{1\over\Delta_0^2}\left(m_b-q\cdot v\right) m_b^2 E_b,\cr
\noalign{\smallskip}
S_8^{(1)} & = {1\over2\Delta_0} \left(m_b E_b -
\bfrac23m_bK_b\right) -
{1\over\Delta_0^2}\left(m_b-q\cdot v\right) m_b^2 E_b,\cr
\noalign{\smallskip}
S_9^{(1)} & = {1\over\Delta_0^2}\left(m_b-q\cdot v\right) m_b E_b.\cr
}}

\subsec{The Order $k^2$ Terms}

The order $k^2$ terms in eq.~\IIIii\ are
\eqn\IIIxxxii{\eqalign{
&-{ 2 k\cdot (m_bv-q)\over \Delta_0^2}\bar u\,
\Bigl\{k^\mu \gamma^\nu + k^\nu \gamma^\mu
-g^{\mu\nu}\slash k - i \epsilon^{\mu\nu\alpha\beta} k_\alpha\gamma_\beta
\Bigr\}\, P_L\, u\cr
&+\left[4{\left(k\cdot\left(m_b v - q\right)\right)^2
\over \Delta_0^3}- {k^2\over \Delta_0^2}\right]
 \bar u\,\Bigl\{ \left(m_b v - q\right)^\mu
\gamma^\nu + \left(m_b v - q\right)^\mu \gamma^\nu -  \left(m_b \slash v
- \slash q\right) g^{\mu\nu}\cr
&\qquad\qquad\qquad\qquad\qquad\qquad\qquad - i \epsilon^{\mu\nu\alpha\beta}
\left(m_b v - q\right)_\alpha \gamma_\beta \Bigr\} P_L\, u.
}}
The matrix element of eq.~\IIIxxxii\ between unpolarized hadrons $H_b$ can be
written in terms of the operator $\bar b\, \gamma^\lambda \, iD^{(\alpha}
iD^{\beta)}\, b$. This matrix element is needed to lowest order in
$1/m_b$, so $b$ can be replaced by the heavy quark field $b_v$, and
$\gamma^\lambda$ replaced by $v^\lambda$. The matrix element needed is
\eqn\IIIxxxiii{
{1\over 2J+1} \sum_s \bra{H_b(v,s)}\bar b_v\, iD^{(\alpha} iD^{\beta)}\,
b_v\ket{H_b(v,s)} = -{2 m_b^2 K_b\over 3} \left(g^{\alpha\beta} - v^\alpha
v^\beta \right),
}
using eq. (3.21). Substituting eq.~\IIIxxxiii\ into eq.~\IIIxxxii\ gives the
contribution of the $k^2$ terms to $\tmunu$,
\eqn\IIIxxxiv{\eqalign{
T_1^{(2)}&= -\bfrac13 m_b^2 K_b\left(m_b-q\cdot v\right)
\left\{ {4\over\Delta_0^3}\left[q^2-(q\cdot v)^2\right] - {3\over
\Delta_0^2}
\right\},\cr
\noalign{\smallskip}
T_2^{(2)}&= - \bfrac23 m_b^3 K_b
\left\{ {4\over\Delta_0^3}\left[q^2-(q\cdot v)^2\right] - {3\over
\Delta_0^2}
\right\} + \bfrac 4 3 m_b^2K_b {v \cdot q\over\Delta_0^2},\cr
\noalign{\smallskip}
T_3^{(2)}&=-\bfrac13 m_b^2 K_b
\left\{ {4\over\Delta_0^3}\left[q^2-(q\cdot v)^2\right] - {3\over
\Delta_0^2}
\right\}+\bfrac23 m_b^2 K_b{1\over\Delta_0^2}.
}}

The spin-dependent contribution to $\tmunu$ for polarized $\Lambda_b$ states is
given in terms of the the matrix element
\eqn\IIIxxxv{
\bra{\Lambda_b(v,s)} \bar b_v\, \gamma^\lambda\gamma_5\, iD^{(\alpha}
iD^{\beta)}
b_v \ket{\Lambda_b(v,s)} = -\bfrac23 m_b^2 K_b \left(g^{\alpha \beta}-v^\alpha
v^\beta\right) s^\lambda,
}
using eq.~\IIIxxi\ and heavy quark spin-symmetry. This gives the order $k^2$
contribution to the spin-dependent form-factors

\eqn\IIIxxxvi{\eqalign{
S_1^{(2)}&= \bfrac13 m_b^2 K_b
\left\{ {4\over\Delta_0^3}\left[q^2-(q\cdot v)^2\right] - {5\over
\Delta_0^2}\right\},\cr
\noalign{\smallskip}
S_6^{(2)}&= \bfrac13 m_b^3 K_b
\left\{ {4\over\Delta_0^3}\left[q^2-(q\cdot v)^2\right] - {3\over
\Delta_0^2}-{2 q\cdot v\over \Delta_0^2}\right\} ,\cr
\noalign{\smallskip}
S_8^{(2)}&= \bfrac13 m_b^2 K_b
\left\{ -{4 m_b\over\Delta_0^3}\left[q^2-(q\cdot v)^2\right] + {3 m_b\over
\Delta_0^2}+{2 q\cdot v\over \Delta_0^2}\right\},\cr
\noalign{\smallskip}
S_9^{(2)}&= \bfrac13 m_b^2 K_b
\left\{ {4\over\Delta_0^3}\left[q^2-(q\cdot v)^2\right] - {5\over
\Delta_0^2}\right\},\cr
\noalign{\smallskip}
S_2^{(2)} &= S_3^{(2)} = 0.\cr
}}

\subsec{The one gluon matrix element}

\insertfig{\centerline{Figure 3}\medskip \centerline{The one gluon
matrix element.}}
{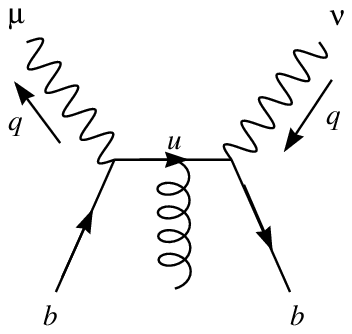}

Finally, one needs to compute the one gluon matrix element of eq.~\IIIi\
given in fig.~3 to determine the
coefficient of the $G^{\alpha\beta}$ operators. The one gluon matrix
element can be expanded in a power series in the momentum $p$ of
the external gluon. The terms of order $p^0$ are identical to the one gluon
matrix element of the operators we have already found, with the gluon field
from
the covariant derivative $D=\partial + i g A$. The terms linear in $p$ are the
matrix element of the operator \eqn\IIIxxxvii{ {g\over 2 \Delta_0^2}\ \bar b\,
G^{\alpha\beta} \epsilon_{\alpha\beta\lambda\sigma}(m_bv-q)^\lambda
\left[g^{\mu\sigma}\gamma^\nu+g^{\nu\sigma}\gamma^\mu -g^{\mu\nu}\gamma^\sigma
+
i \epsilon^{\mu\sigma\nu\tau}\gamma_\tau
\right]\, P_L \,b_v
}
This is a dimension five operator, so we need its matrix element to lowest
order in $1/m_b$. The fields $b$ can be replaced by the heavy quark
fields $b_v$. The matrix elements needed to evaluate the one-gluon
contribution to $\tmunu$ for a spin-averaged hadron $H_b$ are
\eqn\IIIxxxviii{
{1\over 2J+1} \sum_s \bra{H_b (v,s)} \bar b_v\, gG^{\alpha\beta}
\gamma^\lambda \, b_v \ket{H_b (v,s)}, }
and
\eqn\IIIxxxix{
{1\over 2J+1} \sum_s  \bra{H_b (v,s)} \bar b_v\, gG^{\alpha\beta}
\gamma^\lambda \gamma_5\, b_v \ket{H_b (v,s)}.
}
The matrix element in eq.~\IIIxxxviii\ can be simplified by replacing the
 $\gamma$ matrix between $b_v$ fields by $v$. The resultant matrix element must
vanish, because there is no antisymmetric tensor in the indices $\alpha$ and
$\beta$ that can be constructed out of the single four vector $v$. The matrix
element in eq.~\IIIxxxix\ must have the form
\eqn\IIIxl{ {1\over 2J+1} \sum_s \bra{H_b(v,s)} \bar b_v\,
gG^{\alpha\beta} \gamma^\lambda \gamma_5\, b_v \ket{H_b(v,s)} = N
\epsilon^{\alpha\beta\lambda\tau} v_\tau. } The constant $N$ can be evaluated
by contracting both sides of eq.~\IIIxl\ with $\epsilon_{\alpha\beta\lambda
\rho} v^\rho$ to give \eqn\IIIxli{{1\over 2J+1} \sum_s \epsilon_{\alpha\beta
\lambda\rho} v^\rho
\bra{H_b(v,s)} \bar b_v\, gG^{\alpha\beta} \gamma^\lambda \gamma_5\, b_v
\ket{H_b(v,s)} = - 6 N. } Using the spinor identity
$$
\epsilon_{\alpha\beta\lambda\rho}
v^\rho\ \bar b_v\,  \gamma^\lambda \gamma_5\, b_v = -\bar b_v\,
\sigma_{\alpha\beta}\, b_v,
$$
and eq.~\paramdef\ gives
\eqn\IIIxlii{
{1\over 2J+1} \sum_s \bra{H_b(v,s)} \bar b_v\, gG^{\alpha\beta} \gamma^\lambda
\gamma_5\,
 b_v \ket{H_b(v,s)} = \bfrac23 m_b^2 G_b \epsilon^{\alpha\beta\lambda\tau}
v_\tau.
}
Substituting eq.~\IIIxlii\ into \IIIxxxvii\ gives the
contribution of the one gluon operator to $\tmunu$,
\eqn\IIIxliii{\eqalign{
T_1^{(g)} &= -\bfrac13 m_b^2 G_b {m_b-q\cdot v\over \Delta_0^2},\cr
\noalign{\smallskip}
T_2^{(g)} &=  \bfrac23 m_b^2 G_b {m_b\over \Delta_0^2},\cr
\noalign{\smallskip}
T_3^{(g)} &=  -\bfrac13 m_b^2 G_b {1\over \Delta_0^2}.\cr
}}
The gluon terms do not contribute to polarized $\Lambda_b$ decay since
$G^{\alpha\beta}$ has vanishing matrix element at zero recoil between
$\Lambda_b$ states.

\subsec{Summary}

The final expressions for $\tmunu$ and $T_S^{\mu\nu}$ to order $1/m_b^2$ are
obtained by combining
eqs.~\IIIvi, \IIIviii, \IIIxxiv, \IIIxxxi, \IIIxxxiv, \IIIxxxvi, and \IIIxliii,
\eqn\IIIxliv{\eqalign{
T_1 &= {1\over 2\Delta_0}\left(m_b-q\cdot v\right) \left(1+X_b\right)
+ \bfrac23 m_b\left(K_b+G_b\right)\left(-{1\over2\Delta_0} + {q^2-(q\cdot
v)^2\over\Delta_0^2}\right) \cr
&\qquad+ {m_b E_b\over 2 \Delta_0} -
\bfrac13 m_b^2 G_b {m_b-q\cdot v\over \Delta_0^2},
\cr
\noalign{\smallskip}
T_2 &= {m_b\over \Delta_0}\left(1+X_b\right)
+ \bfrac23 m_b\left(K_b+G_b\right)\left({1\over\Delta_0} + {2 m_b q\cdot
v\over\Delta_0^2}\right)
+ {m_b E_b\over \Delta_0} \cr &\qquad + \bfrac43 m_b^2 K_b {q\cdot v\over
\Delta_0^2}+
\bfrac23 m_b^2 G_b {m_b\ \over \Delta_0^2},
\cr
\noalign{\smallskip}
T_3 &= {1\over 2\Delta_0}\left(1+X_b\right)
- \bfrac23 m_b\left(K_b+G_b\right){m_b-q\cdot v\over\Delta_0^2}
\cr &\qquad + \bfrac23 m_b^2 K_b {1\over\Delta_0^2}-
\bfrac13 m_b^2 G_b {1\over \Delta_0^2},\cr
\noalign{\smallskip}
S_1 &= {1\over\Delta_0^3}\left[\bfrac43 m_b^2 K_b \left[q^2-(q\cdot v)^2\right]
\right] - {5 m_b K_b q\cdot v \over 3\Delta_0^2} -\bfrac12(1+\epsilon_b)
{1\over\Delta_0},\cr
\noalign{\smallskip}
S_2 &= {4\over 3 \Delta_0^2} m_b^2 K_b, \cr
\noalign{\smallskip}
S_3 &= {2\over 3 \Delta_0^2} m_b K_b, \cr
\noalign{\smallskip}
S_6&= {1\over\Delta_0^3}\left[\bfrac43 m_b^3 K_b \left[q^2-(q\cdot v)^2\right]
\right] - {5 m_b^2 K_b q\cdot v \over 3\Delta_0^2} -
{5 m_b K_b\over 6 \Delta_0} -\bfrac12(1+\epsilon_b) {m_b\over\Delta_0},\cr
\noalign{\smallskip}
S_8&= -{1\over\Delta_0^3}\left[\bfrac43 m_b^3 K_b \left[q^2-(q\cdot v)^2\right]
\right] + {5 m_b^2 K_b q\cdot v \over 3\Delta_0^2} +
{m_b K_b\over 6 \Delta_0} + \bfrac12(1+\epsilon_b) {m_b\over\Delta_0},\cr
\noalign{\smallskip}
S_9&= {1\over\Delta_0^3}\left[\bfrac43 m_b^2 K_b \left[q^2-(q\cdot v)^2\right]
\right] - {K_b\over \Delta_0^2}\left[m_b q\cdot v + \bfrac 2 3 m_b^2\right] -
\bfrac12(1+\epsilon_b) {1\over\Delta_0},\cr
}}
where
\eqn\IIIxlv{
X_b = - 2 {m_b E_b\over \Delta_0} \left(m_b-q\cdot v\right)
- \bfrac 83 m_b^2{K_b\over \Delta_0^2} \left[q^2 - \left(q\cdot v\right)^2
\right]
+ 2 m_b^2 {K_b\over \Delta_0} .
}

The expressions in eq.~\IIIxliv\ can be simplified using the identity
$E_b=K_b+G_b$. There is an additional simplification in $\Lambda_b$ decay,
where $G_b=0$.

\newsec{Time Ordered Products and the Equations of Motion}

In the above computation, we used the equation of motion to order
$1/m_b$. Another method commonly used in the literature is to treat the
$1/m_b$ terms in the Lagrangian eq.~\IIIxi\ as a
perturbation, so that the equations of motion is $(v \cdot D)\,b_v=0$. The
$1/m_b$ terms in the Lagrangian then give terms that are time-ordered products
of
operators with $1/m_b$ terms in the Lagrangian.
For example the matrix element
$\bra{H_b(v,s)} \bar b_v   i D^\tau b_v \ket{H_b(v,s)}$ (that was needed
in sec.~3,
see eq.~\IIIxix) is zero using $(v \cdot D)\,b_v=0$. However, one now gets an
additional contribution \eqn\IVi{
\bra{H_b(v,s)} T\left\{ \bar b_v   i D^\tau b_v(x)  \ i\int d^4 y
{\CL_1(y)\over{2m_b}}\right\}\ket{H_b(v,s)}, }
where the $1/m_b$ terms in the Lagrangian are denoted by $\CL_1/{2m_b}$,
using the notation of ref.~\ref\falkneubert{A.~Falk and M.~Neubert,
\physrev{D47}{1993}{2965}}.
The matrix element eq.~\IVi\ is equal to $A v^\tau$, where $A$ is the
constant of proportionality. Contracting both sides with $v^\tau$ gives
\eqn\IVii{
A=\bra{H_b(v,s)} T\left\{ \bar b_v   (i v\cdot D) b_v(x)  \ i\int d^4 y
{\CL_1(y)\over{2m_b}}\right\}\ket{H_b(v,s)}.}
$\CL_1(y)$ is a sum of operators of the form $\bar b_v(y) X b_v(y)$, so that
the time ordered product in eq.~\IVii\ is
\eqn\IViii{
\int dy\ T\left\{\bar b_v   (i v\cdot D) b_v(x)\ \bar b_v X b_v(y)\right\}
=\int dy\ T\left\{\bar b_v   (i v\cdot D) iS(x,y) X b_v(y)\right\},
}
where $S(x,y)$ is the b-quark propagator, which satisfies the Green function
equation $(iv\cdot D) S(x,y)=\delta(x,y)$. This converts the T-product into the
local operator $\bar b_v X b_v$, so that $A$ is given by
\eqn\IViv{
A = -\bra{H_b(v,s)}
{\CL_1(y)\over{2m_b}}\ket{H_b(v,s)}, }
the same answer as that obtained by using the equations of motion to order
$1/m_b$.

\newsec{Decay Distributions for Hadrons Containing a b-Quark}

The amplitude $\wmunu$ can be determined by performing a contour
integral of $\tmunu$ from eq.~\IIIxliv\ around the contour $C$ of
fig.~1. This contour integral is trivial to do, and is equivalent to
taking the imaginary part in $\tmunu$ directly by making the replacements
\eqn\Vi{\eqalign{
{1\over \Delta_0} &\rightarrow \delta\left((m_b v - q  )^2 - m^2 \right)\cr
{1\over \Delta_0^2} &\rightarrow -\delta^\prime\left((m_b v - q  )^2 -
m^2 \right)\cr
{1\over \Delta_0^3} &\rightarrow \half\delta^{\prime\prime}
\left((m_b v - q  )^2 - m^2 \right)\cr
}}
in eq.~\IIIxliv. This is very different from the analogous
calculation in deep inelastic scattering. In our
problem, there are non-trivial function of $q$ and $v$, multiplying a few
operators, so that the amplitude has the form $f_i(q^2,q\cdot v) \vev
O_i$. The QCD corrections multiply each operator by an
anomalous dimension, $f_i(q^2,q\cdot v) \vev O_i\rightarrow \lambda_i
f_i(q^2,q\cdot v) \vev O_i$, so the imaginary part of $\tmunu$ is
related directly to the imaginary part of $f_i$.
In deep inelastic scattering, the
different powers of $q\cdot v$ in the expansion of $f_i$ would each be
multiplied by the anomalous dimension of a different twist two operator, so
that the imaginary part of $\tmunu$
is related to the imaginary part of $f_i$ by a non-trivial convolution.

One can now compute the inclusive lepton spectrum by substituting
eq.~\IIIxliv\ and eq.~\Vi\ into eq.~\IIvi, and integrating
over $E_\nu$. The kinematic region is determined by the mass of the
decaying hadron $M_{H_b}$. The $\delta$ functions in $W_i$ restrict the
integration to that given by the parton model kinematics determined by
the quark mass $m_b$. In QCD, one can prove~\ref\guralnik{Z. Guralnik
and A. Manohar, \pl{302}{1993}{103}} that $M_{H_b}-m_b \equiv\bar\Lambda >
0$, so that the parton
model kinematic region is contained within the hadron kinematic region.
Thus there is no dependence on the hadron mass $M_{H_b}$ through the limits
on the region of integration, and the lepton spectrum is determined only
in terms of the quark mass $m_b$. {\it This would not be the case if $\bar
\Lambda$ were negative.} Evaluating the $E_\nu$ integral gives the decay
distribution for $H_b\rightarrow X e \bar\nu_e$,
\eqn\Vii{\eqalign{
&{1\over \Gamma_b}{d\Gamma\over dy\, d\hat q^2} = \theta(z) \biggl\{
12 (y-\hat q^2)(1+\hat q^2-\rho-y)\cr
&+ 12 E_b (2\hat q^4-2\hat q^2\rho+y-2\hat q^2y+\rho y)
+ 8 K_b (2 \hat q^2 - \hat q^4 +\hat q^2\rho-3y)\cr
&\quad + 8 G_b (-\hat q^2+2\hat q^4-2\hat q^2\rho-2y-2\hat q^2y+\rho
y)\biggr\}\cr &+\delta(z){1\over y^2} \biggl\{12E_b \hat q^2 (y-\hat
q^2) (-\hat q^2+2y-y^2)\cr
&\quad+4 K_b(-\hat q^6+9\hat q^4y-6\hat q^2y^2-2\hat q^4y^2-\hat
q^2y^4+y^5)\cr
&\quad+8 \hat q^2G_b (y-\hat q^2)(-\hat q^2+y+y^2)\biggr\}
+K_b\delta^\prime(z) {4 \hat q^2\over y^3} (y^2-\hat q^2)^2(y-\hat q^2)
}}
where
\eqn\Viii{
z = 1+\hat q^2-\rho - {\hat q^2\over y}-y,
}
\eqn\Viv{
\hat q^2 = {q^2\over m_b^2},
\qquad
\rho = {m_j^2\over m_b^2},
\qquad
y = {2 E_e\over m_b},
}
and
\eqn\Vv{
\Gamma_b = \abs{V_{jb}}^2  {m_b^5\over 192\pi^3}.
}
The curve $z=0$ is an edge of the the Dalitz plot in the $\hat q^2-y$ plane.
Integrating eq.~\Vii\ with respect to $\hat q^2$ gives the inclusive lepton
spectrum,
\eqn\Vvi{\eqalign{
&{1\over \Gamma_b}{d\Gamma\over dy } =
\biggl\{2(3-2y)y^2 - 6 y^2 \rho -{ 6y^2\rho^2\over (1-y)^2}
+{ 2(3-y)y^2\rho^3\over (1-y)^3}\biggr\}\cr
&\quad+E_b \biggl\{ 4(3-y) y^2 + {12 y^2 \rho^2 \over (1-y)^3 } - {4
(6-4y+y^2) y^2\rho^3\over (1-y)^4}\biggr\}\cr
&\quad+K_b\biggl\{-{4 y^2
(9+2y)\over 3} + {4 y^2 (2y^2-2y-3)\rho^2\over(1-y)^4} + {4y^2 (18-10y
+5y^2-y^3)\rho^3\over 3(1-y)^5}\biggr\}\cr
&\quad+G_b\biggl\{-{4 y^2
(15+2y)\over 3} + {8 y^2 (3-2y)\rho\over(1-y)^2} + {12y^2 \rho^2\over
(1-y)^2} + {8y^2 (-6+4y-y^2)\rho^3\over 3(1-y)^4}\biggr\}.\cr
}}
This agrees with the result of Bigi \etal.
Integrating eq.~\Vvi\ with respect to $y$ gives the total decay rate
\eqn\Vvii{\eqalign{
{1\over\Gamma_b}\Gamma &= \biggl\{1-8\rho+8\rho^3-\rho^4-12\rho^2\log\rho
\biggr\}\cr
&\quad+E_b\biggl\{5-24\rho+24\rho^2-8\rho^3+3\rho^4-12\rho^2\log\rho\biggr\}\cr
&\quad+K_b\biggl\{-6+32\rho-24\rho^2-2\rho^4+24\rho^2\log\rho\biggr\}\cr
&\quad+G_b\biggl\{-2+16\rho-16\rho^3+2\rho^4+24\rho^2\log\rho\biggr\},
}}
which also agrees with Bigi \etal.

The decay rate for $b\rightarrow u$ is given by eqs.~\Vii--\Vvii\ in the limit
that $\rho\rightarrow 0$. This limit must be taken carefully because of the
$1/(1-y)$ singularities in eq.~\Vvi. The resulting expressions are
\eqn\Vviii{\eqalign{
&{1\over \Gamma_b}{d\Gamma\over dy \,d\hat q^2} = \theta(z) \biggl\{
12 (y-\hat q^2)(1+\hat q^2-y)
+ 12 E_b (2\hat q^4+y-2\hat q^2y)\cr
& + 8 K_b (2 \hat q^2 - \hat q^4-3y)+ 8 G_b (-\hat q^2+2\hat q^4-2y-2\hat
q^2y)\biggr\}\cr
&+\delta(z){1\over y^2} \biggl\{12E_b \hat q^2 (y-\hat q^2)(-\hat
q^2+2y-y^2)\cr &\quad+4 K_b(-\hat q^6+9\hat q^4y-6\hat q^2y^2-2\hat q^4y^2-\hat
q^2y^4+y^5)\cr &\quad+8 \hat q^2G_b (y-\hat q^2)(-\hat
q^2+y+y^2)\biggr\}+\delta^\prime(z) K_b {4 \hat q^2\over y^3} (y^2-\hat
q^2)^2(y-\hat q^2),
}}
where now
\eqn\Vxix{
z = 1+\hat q^2 - {\hat q^2\over y}-y,
}
\eqn\Vx{\eqalign{
{1\over \Gamma_0}{d\Gamma\over dy } &=
\biggl\{2(3-2y)y^2\biggr\}+E_b \biggl\{ 4(3-y) y^2 + 2 \delta(1-y)\biggr\}\cr
&\quad+K_b\biggl\{-{4 y^2
(9+2y)\over 3}-\bfrac43 \delta(1-y) + \bfrac23 \delta^\prime(1-y) \biggr\}\cr
&\quad+G_b\biggl\{-{4 y^2
(15+2y)\over 3} + \bfrac{16}{3} \delta(1-y) \biggr\},\cr
}}
and
\eqn\Vxi{
\Gamma = \Gamma_b\left( 1+
5E_b-6K_b-2G_b\right).
}
The results eqs.~\Vx\ and \Vxi\ agree with Bigi \etal.
The $\delta^\prime$-function in eq.~\Vx\ is present because the parton
model decay
distribution $\propto 2(3-2y)y^2$ does not vanish at the end point. This will
be explained in more detail in the next section.

The decay distributions for a polarized $\Lambda_b$ have the form $A + B
\cos\theta$. The coefficient $A$ is half the value of the corresponding
decay distribution for an unpolarized $\Lambda_b$ given in
eqs.~\Vii--\Vxi. The results can be simplified for $\Lambda_b$ by
setting $E_b=K_b$ and $G_b=0$. The coefficients of the $\cos\theta$
terms are
\eqn\Vxii{\eqalign{
&{1\over \Gamma_b}{d\Gamma\over dy\, d\hat q^2\,d\!\cos\theta} =\ldots
+\Biggl[{\theta(z)\over y} \biggl\{
6\left(1+\epsilon_b\right)
(y-\hat q^2)(-2\hat q^2+y+\hat q^2 y-\rho y - y^2)\cr
& + 2 K_b (-6 \hat q^4 +12 \hat q^2 y + 4\hat q^4 y -
4  \hat q^2 \rho y-3y^2-6\hat q^2 y^2 + 3 \rho y^2)
\biggr\}\cr
&+\delta(z){2\over y^2}
\biggl\{
K_b(-4\hat q^6+2\hat q^4y+3\hat q^4y^2-\hat q^2y^3-\hat
q^2y^4+y^5)\biggr\}\cr
&-K_b\delta^\prime(z) {2 \hat q^2\over y^3} (y^2-\hat q^2)^2(y-\hat q^2)
\Biggr]\cos\theta,
}}
\eqn\Vxiii{\eqalign{
&{1\over \Gamma_b}{d\Gamma\over dy\,d\!\cos\theta } =\ldots +\Biggl[
\left(1+\epsilon_b\right)
\biggl\{(1-2y)y^2 - 3 y^2 \rho +{ 3y^2\rho^2\over (1-y)^2}
-{ (1+y)y^2\rho^3\over (1-y)^3}\biggr\}\cr
&\quad+K_b\biggl\{-{10 y^3
\over 3} + {2 y^3 (5-2y)\rho^2\over(1-y)^4} - {4y^3 (5+2y-y^2)
\rho^3\over 3(1-y)^5}\biggr\}\Biggr]\cos\theta,
}}
\eqn\Vxiv{\eqalign{
{1\over\Gamma_b}{d\Gamma\over
d\!\cos\theta} &=\ldots +\left(1+\epsilon_b-K_b\right)\Biggl[
\biggl\{-\bfrac16+2\rho+6\rho^2-\bfrac{22}{3}\rho^3-
\bfrac12\rho^4\cr
&\qquad+6\rho^2\log\rho+4\rho^3\log\rho
\biggr\}\Biggr]\cos\theta.
}}

The $\rho\rightarrow0$ limits of the polarized $\Lambda_b$ distributions are
\eqn\Vxv{\eqalign{
&{1\over \Gamma_b}{d\Gamma\over dy\, d\hat q^2\,d\!\cos\theta} =\ldots
+\Biggl[{\theta(z)\over y} \biggl\{
6\left(1+\epsilon_b\right)
(y-\hat q^2)(-2\hat q^2+y+\hat q^2 y- y^2)\cr
& + 2 K_b (-6 \hat q^4 +12 \hat q^2 y + 4\hat q^4 y -3y^2-6\hat q^2 y^2 )
\biggr\}\cr
&+\delta(z){2\over y^2}
\biggl\{
K_b(-4\hat q^6+2\hat q^4y+3\hat q^4y^2-\hat q^2y^3-\hat
q^2y^4+y^5)\biggr\}\cr
&-K_b\delta^\prime(z) {2 \hat q^2\over y^3} (y^2-\hat q^2)^2(y-\hat q^2)
\Biggr]\cos\theta,
}}
\eqn\Vxvi{\eqalign{
&{1\over \Gamma_b}{d\Gamma\over dy\,d\!\cos\theta } =\ldots +\Biggl[
\left(1+\epsilon_b\right)
\biggl\{(1-2y)y^2  \biggr\}\cr
&\quad+K_b\biggl\{-{10 y^3
\over 3} +\delta(1-y)-\bfrac13\delta^\prime(1-y)\biggr\}\Biggr]\cos\theta,
}}
\eqn\Vxvii{
{1\over\Gamma_b}{d\Gamma\over
d\!\cos\theta} =\ldots -\bfrac16\left(1+\epsilon_b-K_b\right)\cos\theta.
}

The formul\ae\ obtained here will be discussed in detail in the following
sections.

\newsec{Physical Interpretation of the $1/m_b^2$ Corrections}

There is a way to obtain most of the $1/m_b^2$ corrections which also
provides a  physical picture of these corrections.
One can obtain a class of $1/m_b^2$ corrections by taking the lowest order
expression eq.~\IIIvi\ for $\tmunu$ (or $\wmunu$) and smearing it over a
distribution of b-quark momenta in the $H_b$ hadron. This will give the terms
proportional to $E_b$ and $K_b$ in secs.~4--5, but not those proportional to
$G_b$. This is why we did not simplify the results using $E_b=G_b+K_b$.

We begin with the spin independent terms.  The b-quark momentum in
$\ket{H_b(v,s)}$ can be written as $p=m_b v + k$. The lowest
order (parton model) expressions  $\tmunu_0(q,v,m_b,m_j)$ were obtained
by considering the decay of a quark of mass $m_b$ and velocity $v$ in the rest
frame $v=(1,0,0,0)$. A quark with momentum $m_b v + k$ can be considered to be
an
on-shell quark with mass $m_b^\prime$ and velocity $v^\prime$, where $v^{\prime
\,2}=1$, and $m_b^\prime v^\prime = m_b v + k$. The decay rate of such a quark
can be obtained by using  $\tmunu_0(q,
v^\prime,m_b^\prime,m_j)/v^{\prime\, 0}$, where $v^{\prime \, 0}$ is the
Lorentz
time dilation factor for a moving particle. The value of $\tmunu$ is obtained
by
averaging $\tmunu_0(q,v^\prime,m_b^\prime,m_j)/v^{\prime\,0}$ over a
distribution of b-quark momenta. This average is  most easily done by writing
$v^{\prime\,0}=v\cdot m_b^\prime v^{\prime}/ m_b^\prime$.
$m_b^\prime\tmunu_0(q,
v^\prime,m_b^\prime,m_j) / v \cdot m_b^\prime v^\prime$ can be written as a
function of $q$, $m_j$ and the product $m_b^\prime v^\prime$. This makes
the averaging simple, because one can use the substitution
$m_b^\prime v^\prime = m_b v + k$ to rewrite the expression in terms of
$m_b v$ and $k$, and then averaging over $k$. Since $q$ is unaffected by the
averaging, terms proportional to $q^\mu$ or $q^\nu$ will remain
proportional to $q^\mu$ and $q^\nu$. Thus averaging $T_4$ or $T_5$
will not produce $T_{1-3}$. Similarly, averaging $T_1$ or $T_3$ will
produces terms only proportional to $T_1$ or $T_3$, but averaging $T_2$
produces terms proportional to both $T_2$ and $T_1$.

As a simple example, we will consider the average of the piece of
$T^{\mu\nu}$ containing $T_3$ explicitly.
The average we need is
\eqn\VIi{\eqalign{
&\vev{{m_b^\prime \over v \cdot m_b^\prime v^\prime}\left(
-i\epsilon^{\mu\nu\alpha\beta}v_\alpha^\prime q_\beta\ T_3(q,q\cdot
v^\prime,m_b^\prime)\right)} \cr &=
\vev{
{1 \over v \cdot m_b^\prime v^\prime}
\left( -i\epsilon^{\mu\nu\alpha\beta}m_b^\prime v_\alpha^\prime q_\beta
{1\over
(m_b^\prime v^\prime -q )^2 - m_j^2}\right)}\cr
&= \vev{{1 \over v \cdot (m_b v + k)}\left(
-i\epsilon^{\mu\nu\alpha\beta}(m_b v + k)_\alpha q_\beta {1\over
(m_b v +k  -q )^2 - m^2_j}\right)}
}}
The averages over $k$ can be written in terms of
\eqn\VIii{\eqalign{
&\vev {k^\alpha}  = E_b m_b v^\alpha ,\cr
&\vev{k^\alpha k^\beta} = \frac13\vev{k^2} \left(g^{\alpha\beta} -
v^\alpha v^\beta\right) =  -\frac23 K_b m_b^2 \left(g^{\alpha\beta} -
v^\alpha v^\beta\right),
}}
where $E_b$ and $K_b$ are the mean total energy and mean kinetic energy
in units of $m_b^2$. This gives the terms in eq.~\IIIxliv\ with the
exception of the $G_b$ term. A similar computation also reproduces $T_1$
and $T_2$ in eq.~\IIIxliv.

The above averaging procedure computes the $1/m_b^2$ corrections using a
distribution
of quark momenta $k$ in a hadron $H_b$. Here $k$ is considered to be a
number, not an operator. The operator product expansion gives a similar
result, with $k$
replaced by the covariant derivative $iD$. The commutator of two covariant
derivatives is proportional to the gluon field-strength, $[D^\alpha,D^\beta]=ig
G^{\alpha\beta}$. The covariant derivatives commute if we neglect all terms
involving $G^{\alpha\beta}$, \ie\ all terms involving $G_b$.  Thus the $1/m_b$
terms obtained by the averaging method are identical to those obtained using
the operator product expansion neglecting $G^{\alpha\beta}$. This result holds
to all orders in the $1/m_b$ expansion due to reparameterization
invariance \reparaminv.

Our averaging procedure provides a useful check on the total decay rate.
The total decay rate (neglecting $1/m_b$ corrections) can be written as
$\Gamma_0=m_b^5 f(m^2/m_b^2)$. The total decay rate for a quark distribution is
given by averaging $$m_b^{\prime\,5} f(m^2/m_b^{\prime\,2})/v^{\prime 0} =
m_b^{\prime\,6} f(m^2/m_b^{\prime\,2})/ v \cdot m_b^\prime v^\prime.$$ Now
$m_b^{\prime\,2} = m_b^{\prime\, 2} v^{\prime\, 2} = (m_b v + k)^2 = m_b^2 (1 +
2E_b -2 K_b)$, and $ v \cdot m_b^\prime v^\prime = m_b (1 + E_b)$. This gives
\eqn\VIiii{ \Gamma = m_b^5 (1+E_b)^{-1} (1+2 E_b -2 K_b)^3 f(\rho (1 + 2 E_b
-2 K_b)^{-1}), }
where $\rho=m^2/m_b^2$.
Expanding this and retaining corrections to order $1/m_b^2$ gives
\eqn\VIiv{
\Gamma = \Gamma_0 + E_b (5\Gamma_0 - 2 \rho {d\Gamma_0\over d\rho} ) +
K_b (-6\Gamma_0 +2  \rho {d\Gamma_0\over d\rho} ) .
}
The result eq.~\Vvii\ agrees with this check.

One can also use the averaging procedure to determine the leptonic spectrum.
Recall that we needed to compute the average of $m_b^\prime T_0^{\mu\nu}/v\cdot
m_b^\prime v^\prime$, where $m_b^\prime T_0^{\mu\nu}$ is a function only of
$m_b^\prime v^\prime$ and $q$, and has mass dimension zero. This can be written
as
\eqn\VIv{\eqalign{
&\vev{ {m_b^\prime T_0^{\mu\nu}\over v\cdot
m_b^\prime v^\prime} } = {1\over m_b (1+E_b)}
\Biggl\langle m_b T_0^{\mu\nu} + k^\alpha {\partial\over \partial m_b v^\alpha}
m_b T_0^{\mu\nu} \cr
&\qquad\qquad+ \bfrac12 k^\alpha k^\beta {\partial^2 \over \partial m_b
v^\alpha \partial m_b v^\beta} m_b T_0^{\mu\nu} \Biggr\rangle,\cr
&=\Biggl[  T_0^{\mu\nu} -  E_b T_0^{\mu\nu}
+ E_b v^\alpha {\partial\over \partial m_b v^\alpha} m_b
T_0^{\mu\nu}\cr
&\qquad\qquad -\bfrac13 K_b  m_b \left(g^{\alpha\beta}-v^\alpha
v^\beta\right)
{\partial^2 \over \partial m_b v^\alpha
\partial m_b v^\beta} m_b T_0^{\mu\nu} \Biggr],
}}
using eq.~\VIii. The decay rate depends on $L_{\mu\nu} T^{\mu\nu}\equiv F$
which has mass dimension two, and is a function only of $k_e$, $k_\nu$, $mv$,
$m_j$ and $q$. One can rewrite $F$ in terms of the dimensionless
variables $y,
\hat q$ $\rho$, and an overall factor of $m_b^2$. The variables $y$, $\hat
q$ and $\rho$ were defined in eq.~\Viv, and $x$ is
defined by
\eqn\VIvi{
x = {2 E_\nu \over m_b} = {2m_b v \cdot k_\nu\over (m_b v)^2}.
}
The averaging formula eq.~\VIv\ for $T^{\mu\nu}$ implies that $F$ can be
written
in terms of the lowest order expression $F_0=L_{\mu\nu} T_0^{\mu\nu}$, by
differentiating with respect to $m_bv$. In performing the
differentiation, it is
important to remember that $T^{\mu\nu}_0$ and $F_0$ are to be considered as
functions only of the product $m_bv$, not of $m_b$ and $v$ separately. Thus
$m_b$ is an implicit function of $v$, with
\eqn\VIci{
{\partial\over \partial m_b v^{\alpha}}m_b ={\partial\over
\partial m_b v^{\alpha}} \left[ (m_b v)\cdot (m_b v)\right]^{1/2}
= v_{\alpha}.
}
Using the partial derivatives
\eqn\VIcii{\eqalign{
{\partial x\over \partial (m_b v^\alpha)}& = \left[{2 k_{\nu\alpha}\over
m_b^2} - {2xv_\alpha\over m_b}\right] ,\cr
{\partial y\over \partial (m_b
v^\alpha)} &= \left[{2k_{e\alpha}\over m_b^2} - {2yv_\alpha\over
m_b}\right],\cr}
\qquad
\eqalign{
{\partial \hat q^2\over \partial (m_b v^\alpha)} &= - {2\hat q^2\over
m_b} v_\alpha,\cr
{\partial\rho\over\partial (m_b v^\alpha)} &= -
{2\rho\over m_b} v^\alpha,\cr
}}
and eq.~\VIv\ our averaging procedure implies that
\eqn\VIvii{\eqalign{
&F = \biggl\{1 + E_b\Bigl[ 1-2 \rho {\partial \over \partial \rho} - 2 \hat q^2
{\partial\over \partial\hat q^2} - y{\partial\over \partial y} -
x{\partial\over \partial x} \Bigr]\cr & \qquad+K_b \Bigl[-2
+2 \rho {\partial \over \partial \rho}+ 2 \hat q^2
{\partial\over \partial\hat q^2} + 2 y{\partial\over \partial y} +2
x{\partial\over \partial x} \cr
&\qquad\qquad+\bfrac13 y^2 {\partial^2\over \partial y^2}
+\bfrac13 x^2 {\partial^2\over \partial x^2}
+\bfrac23 (x y - 2 q^2)  {\partial^2\over \partial x\partial y} \Bigr]
\biggr\} F_0.
}}
The differential decay rate is given by integrating $F$ over the phase space,
which is proportional to $dy\, dx \, d\hat q^2$. Integrating
eq.~\VIvii\ with respect to $x$ gives the formula
\eqn\VIviii{\eqalign{
{d\Gamma\over dy\, d\hat q^2} &=
\biggl\{1 + E_b\Bigl[ 2 -2 \rho {\partial \over \partial \rho}- 2 \hat q^2
{\partial\over \partial\hat q^2} - y{\partial\over \partial y} \Bigr]\cr
&\qquad +K_b \Bigl[-\bfrac{10}{3} +2 \rho {\partial \over \partial
\rho}+ 2 \hat
q^2 {\partial\over \partial\hat q^2} + \bfrac43 y{\partial\over \partial y}
+\bfrac13 y^2 {\partial^2\over \partial y^2}\Bigr]
\biggr\}{d\Gamma_0\over dy\, d\hat q^2},
}}
where $d\Gamma_0/dy\,d\hat q^2$ is the parton model decay rate obtained by
setting $E_b,K_b,G_b\rightarrow 0$ in eq.~\Vii.
Integrating eq.~\VIviii\ with
respect to $\hat q^2$ gives \eqn\VIix{\eqalign{
{d\Gamma\over dy} &=
\biggl\{1 + E_b\Bigl[ 4-2 \rho {\partial \over \partial \rho} - y{\partial\over
\partial y} \Bigr]\cr &\qquad +K_b \Bigl[-\bfrac{16}{3} +2 \rho
 {\partial \over \partial \rho} + {4\over 3} y{\partial\over
\partial y}  +\bfrac13 y^2 {\partial^2\over \partial y^2}\Bigr]
\biggr\} {d\Gamma_0\over dy}.
}}
Integrating eq.~\VIix\ with respect to $y$ gives
\eqn\VIx{
{\Gamma} =
\biggl\{1 + E_b\Bigl[ 5-2 \rho {\partial \over \partial \rho}
\Bigr]
+K_b \Bigl[-6 +2 \rho {\partial \over \partial \rho}
 \Bigr] \biggr\} {\Gamma}_0,
}
which is the same result we obtained in eq.~\VIiv.
Eqs.~\VIviii--\VIx\ agree with eqs.~\Vii--\Vvii\ on setting $G_b\rightarrow 0$.
One can also understand the origin of the $\delta$-function terms in sec.~5.
Since the decay distribution does not vanish at the end point, the derivatives
in eq.~\VIix\ produce $\delta$-functions and derivatives of
$\delta$-functions.

One can also apply the averaging method to obtain the spin-dependent
form-factors. One considers the b-quark in the hadron to have a distribution of
spin and momentum, with
\eqn\VIxi{\eqalign{
&\vev{S^\mu} = (1+\epsilon_b+K_b) s^\mu,\cr
&\vev{S^\mu k^\nu} = m_b K_b s^\mu v^\nu + \bfrac23 m_b K_b s^\nu v^\mu,\cr
&\vev{S^\mu k^\alpha k^\beta} = - \bfrac23 m_b^2 K_b (g^{\alpha\beta}-v^\alpha
v^\beta ) s^\mu
}}
where $S^\mu$ is the quark spin, and $s^\mu$ is the hadron spin.

\newsec{Decay Distributions for Hadrons Containing a c-Quark}

The decay distributions for semileptonic c-quark decay can be readily obtained
from the calculations in the previous section for b-quark decay. The charged
lepton distribution in c-decay is equal to the neutrino distribution in
b-decay, and vice-versa. This is equivalent to changing the signs of $W_3$,
$G_3$, $G_8$ and $G_9$. The lepton energy spectra are less singular than
for b-decay, because the free quark decay rate vanishes at the endpoint.
The double differential c-decay distribution for an unpolarized hadron $H_c$
containing a c-quark is
\eqn\VIIi{\eqalign{
&{1\over \Gamma_c}{d\Gamma\over dy\, d\hat q^2} =\theta(z) \biggl\{
12y (1-\rho -y) + 12 y E_c (1+\rho)- 24 y K_c + 8y G_c (\rho  - 2 )\biggr\}\cr
&+\delta(z){1\over y} \biggl\{
12 E_c \hat q^2 (y-1)(\hat q^2-2y+y^2)\cr
&\quad+4 K_c (3 \hat q^4 - 6 \hat q^2 y - 4 \hat q^4 y + 6 \hat q^2 y^2
+y^4)\cr
&\quad+ 8 \hat q^2 G_c (1-y)(\hat q^2+y-y^2)\biggr\}+\delta^\prime(z)K_c
{4 \hat q^2\over y^2} (y^2-\hat q^2)^2(1-y),
}}
where
\eqn\VIIii{\Gamma_c = \abs{V_{jc}}^2 {m_c^5\over 192\pi^3}.}
Integrating this with respect to $\hat q^2$ and then $y$ gives
\eqn\VIIiii{\eqalign{
{1\over \Gamma_c}{d\Gamma\over dy } &=
\biggl\{12(1-y)y^2- 24 y^2 \rho + { 12 y^2\rho^2\over (1-y)}\biggr\}\cr
&\quad+E_c \biggl\{ 12(2-y) y^2 + {12 y^2 (y-2) \rho^2 \over (1-y)^2
}\biggr\}\cr &\quad+K_c\biggl\{-{8 y^2
(3+y)} + {8 y^2 (3-2y)\rho^2\over(1-y)^3} \biggr\}\cr
&\quad+G_c\biggl\{-{8 y^3} + {8 y^3\rho^2\over(1-y)^2}\biggr\},
}}
and
\eqn\VIIiv{\eqalign{
{1\over\Gamma_c}\Gamma &= \biggl\{1-8\rho+8\rho^3-\rho^4-12\rho^2\log\rho
\biggr\}\cr
&\quad+E_c\biggl\{5-24\rho+24\rho^2-8\rho^3+3\rho^4-12\rho^2\log\rho\biggr\}\cr
&\quad+K_c\biggl\{-6+32\rho-24\rho^2-2\rho^4+24\rho^2\log\rho\biggr\}\cr
&\quad+G_c\biggl\{-2+16\rho-16\rho^3+2\rho^4+24\rho^2\log\rho\biggr\}.
}}
The $\rho\rightarrow 0$ limits of eqs.~\VIIi---\VIIiv\ are
\eqn\VIIv{\eqalign{
&{1\over \Gamma_c}{d\Gamma\over dy\, d\hat q^2} =\theta(z) \biggl\{
12 (y -y^2) + 12 y E_c - 24 y K_c - 16 y G_c \biggr\}\cr
&\qquad+\delta(z){1\over y} \biggl\{
12 E_c\hat q^2 (y-1)(\hat q^2-2y+y^2) \cr
&\qquad
-4 K_c (-3 \hat q^4 + 6 \hat q^2 y + 4 \hat q^4 y - 6 \hat q^2 y^2 -y^4)\cr
&- 8 \hat q^2 G_c (y-1)(\hat q^2+y-y^2)\biggr\}+\delta^\prime(z)K_c {4 \hat
q^2\over y^3} (y^2-\hat q^2)^2(1-y),
}}
\eqn\VIIvi{\eqalign{
{1\over \Gamma_c}{d\Gamma\over dy } &=
\biggl\{12(1-y)y^2\biggr\}+E_c \biggl\{ 12(2-y) y^2\biggr\}\cr
&\qquad+K_c\biggl\{-{8 y^2 (3+y)+4\delta(1-y)}\biggr\}+G_c\biggl\{-{8 y^3}
\biggr\}, }}
\eqn\VIIvii{
\Gamma = \Gamma_0\left\{1+ 5 E_c-6 K_c - 2 G_c\right\}.
}
The $\cos\theta$ terms in the decay distributions for a polarized $\Lambda_c$
are
\eqn\VIIviii{\eqalign{
&{1\over \Gamma_c}{d\Gamma\over dy\, d\hat q^2\,d\!\cos\theta} =\ldots+\Biggl[
\theta(z)\left(1+\epsilon_c\right) \biggl\{
 6 y (1-\rho -y)\cr
&- 6 y(1-\rho)  K_c \biggr\}
+\delta(z) \biggl\{
2 K_c (-\hat q^4 - 3 \hat q^2 y  + 3 \hat q^2 y^2 +y^3)
\biggr\}\cr
&\qquad+\delta^\prime(z)K_c {2 \hat q^2\over y^2} (y^2-\hat
q^2)^2(1-y)\Biggr]\cos\theta,
}}
\eqn\VIIix{\eqalign{
&{1\over \Gamma_c}{d\Gamma\over dy\,d\!\cos\theta } =\ldots+\Biggl[
\left(1+\epsilon_c\right)\left\{
6(1-y)y^2- 12 y^2 \rho + { 6 y^2\rho^2\over (1-y)}\right\}\cr
&\qquad\qquad+K_c\left\{-{10 y^3} + {2 y^3 (5-3y)\rho^2\over(1-y)^3}
\right\}\Biggr]\cos\theta,
}}
\eqn\VIIx{
{1\over\Gamma_c}{d\Gamma\over
d\!\cos\theta} = \ldots+\left(1+\epsilon_c-K_c\right)\left[
\left\{\bfrac12-4\rho+4\rho^3-\bfrac12\rho^4-6\rho^2\log\rho
\right\}\right]\cos\theta.
}
The $\rho\rightarrow0$ limits of these distributions are
\eqn\VIIxi{\eqalign{
&{1\over \Gamma_c}{d\Gamma\over dy\, d\hat q^2\,d\!\cos\theta} =\ldots+\Biggl[
\theta(z)\left(1+\epsilon_c\right) \biggl\{
 6 y (1-y)\cr
&- 6 y  K_c \biggr\}
+\delta(z) \biggl\{
2 K_c (-\hat q^4 - 3 \hat q^2 y  + 3 \hat q^2 y^2 +y^3)
\biggr\}\cr
&\qquad+\delta^\prime(z)K_c {2 \hat q^2\over y^2} (y^2-\hat
q^2)^2(1-y)\Biggr]\cos\theta,
}}
\eqn\VIIxii{\eqalign{
&{1\over \Gamma_c}{d\Gamma\over dy\,d\!\cos\theta } =\ldots+\Biggl[
\left(1+\epsilon_c\right)\left\{
6(1-y)y^2\right\}\cr
&\quad+K_c\left\{-{10 y^3} + 2\delta(1-y)\right\}\Biggr]\cos\theta,
}}
\eqn\VIIxiii{
{1\over\Gamma_c}{d\Gamma\over
d\!\cos\theta} = \ldots+\bfrac12\left(1+\epsilon_c-K_c\right)\cos\theta.
}

\newsec{Inclusive $H_b \rightarrow X_ue\bar\nu_e$ near the boundary of
phase space}

For an exclusive decay $H_b \rightarrow Xe\bar\nu_e$ the kinematically
allowed region of phase space is
\eqn\IXi{0 < q^2 < 2E_e M_{H_b}+ {2M^2_X E_e\over (2E_e - M_{H_b})}.}
The maximum value of the electron energy $E_e^{(\max)}$ occurs when the
right hand side of eq.~\IXi\ is zero,
\eqn\IXii{E_e^{(\max)} = {M_{H_b}^2 - M_X^2\over 2M_{H_b}}.}
The results of sec.~5 show that in QCD, the inclusive decay kinematics are
governed by the quark mass $m_b$, rather than the hadron mass $M_{H_b}$. The
kinematically allowed region in the Dalitz plot is
\eqn\IXiii{0 < q^2 < 2E_e m_b + {2m_j^2 E_e\over (2E_e - m_b)},}
where $m_j$ is the charm quark mass for $b \rightarrow c$ transitions
and zero for $b \rightarrow u$ transitions (neglecting light quark masses).
The difference between these two kinematic regions is shown in fig.~4
for $b\rightarrow u$ decay.

\insertfig{\centerline{Figure 4}\medskip
The allowed region of the Dalitz plot for
$B\rightarrow X_u e\bar \nu_e$ decay. $E_e$ is in GeV, and $q^2$ is in
GeV${}^2$. The outer curve is the region allowed using physical hadron masses.
The inner curve is obtained using quark quark masses, with $\bar\Lambda$ for
the $B$ meson chosen to be 500~MeV. The free quark decay distribution is
non-zero only inside the inner triangle. }{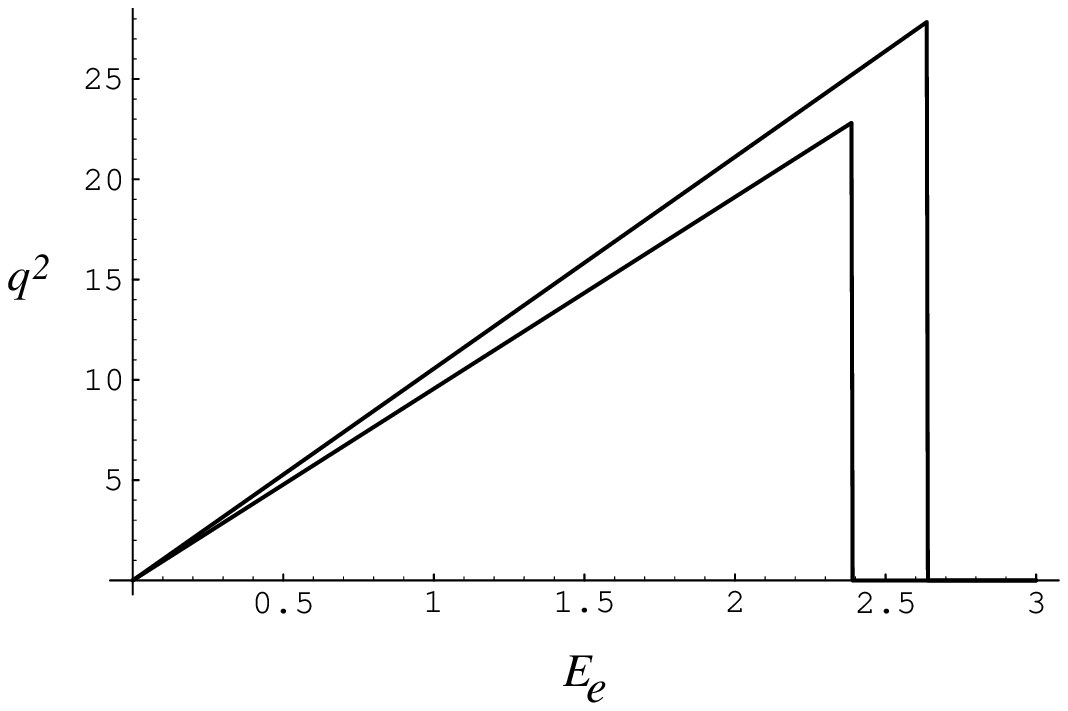}

For $b \rightarrow u$ transitions the region \IXiii\
of the $q^2, E_e$ plane becomes
the interior of a right triangle with sides $q^2 = 0, q^2 = 2E_e m_b$
and $E_e = m_b/2$.  Since $m_b$ is less than $M_{H_b}$ a comparison of
eqs.~\IXiii\ and \IXi\
reveals that a part of the kinematically allowed phase
space is not populated.  This part corresponds to the production of
states with mass squared
\eqn\IXiv{M_{Xu}^2 \ltap (M_{H_b} - m_b) M_{H_b} \left(1 - {q^2\over
M_{H_b}^2}\right).}
The physical origin of this discrepancy has been discussed by Isgur,
Scora, Grinstein and Wise~\ref\isgw{N.~Isgur, D.~Scora, B.~Grinstein and
M.B.~Wise, \physrev{D39}{1989}{799}}\
and by Isgur~\ref\isgur{N.~Isgur, \physrev{D47}{1993}{2782}}.  For simplicity,
consider QCD in
the large $N_c$ limit where the final state $X$ must be a $\bar qq$ bound
state (nonresonant final states are produced with an amplitude
suppressed by a factor of $1/\sqrt{N_c}$).  Provided $q^2/M_{H_b}^2$ is not
very close to unity, the production of states with masses much less than
in eq.~\IXiv\ is strongly
suppressed by hadronic form factors.  For very large $m_b$ this form
factor suppression arises from the transfer of a large momentum to the
spectator antiquark by a single hard gluon.  Clearly such effects are
subdominant to those given by eq.~\Vviii.  However, when the mass of the
final $\bar qq$ resonance becomes of order $\sqrt{m_b \lqcd}$
(see eq.~\IXiv), the antiquark in the $X$ meson has such a broad
distribution of momentum that no form factor suppression is required.
Thus the differential decay rate in eq.~\Vviii\ corresponds to a sum over
exclusive final states with masses greater than $\sim \sqrt{m_b
\lqcd}$.  Note that for large $m_b$ these hadronic masses are
much greater than the QCD scale.

There are theoretical limitations on the extent to which our prediction
for the differential $H_b \rightarrow X_u e\bar\nu_e$ decay rate $d\Gamma/dq^2
dE_e$ can be compared with
experiment.  In the region of phase space very near $q^2 = M_{H_b}^2$, which
corresponds to low mass final hadronic states recoiling at low momentum,
our expression for the differential decay rate is not valid (the
operator product expansion cannot be justified for low mass states).  Also,
the appearance in eq.~\Vviii\ of a delta function and its first derivative
indicates that along the boundaries $q^2 = 2m_b E_e$ and $E_e = m_b/2$
the differential decay rate must be smeared over a region of electron
energies.  The amount of smearing necessary is determined by demanding
that corrections proportional to $K_b, E_b$ and $G_b$ give a
contribution to the smeared differential rate that is small compared
with the leading ``free quark decay'' contribution.

For definiteness consider the boundary $q^2 = 2m_b E_e$.  In the free $b
\rightarrow ue\bar\nu_e$ decay it corresponds to a configuration (in the
$b$ rest frame) where the electron and $u$ quark are moving in the same
direction and the
antineutrino goes in the opposite direction.  Since the weak current is
left handed the free quark decay amplitude vanishes here by angular
momentum conservation.  Define (recall $y = 2E_e/m_b$, $\hat q^2 =
q^2/m_b^2$)
\eqn\IXv{S(q^2) = \int dy {1\over \Gamma_b} {d^2\Gamma\over dy d\hat q^2}
W(y),}
where
\eqn\IXvi{W = {1\over\sqrt{\pi} \epsilon} e^{-(y - (\hat q^{2} +
\epsilon))^{2}/\epsilon^{2}}.}
$S(q^2)$ corresponds to smearing the differential cross section at $y =
\hat q^2 + \epsilon$ over a region of $y$ of order $\epsilon$.  (The
averaging is at $y = \hat q^2 + \epsilon$ instead of $y = \hat q^2$ so
that the term containing a derivative of a delta function is not made
artificially small.)
Demanding that the leading free quark decay contribution to $S(q^2)$ be
large compared to that of the corrections proportional to $K_b, E_b$ and
$G_b$ gives the condition
\eqn\IXci{
\epsilon^2 > (\lqcd/m_b)^2.
}
This corresponds to smearing over a range of electron energies $\Delta
E_e > \lqcd$.  In deriving eq.~\IXci\ the matrix elements $K_b,
E_b$ and $G_b$ were estimated to be of order $(\lqcd/m_b)^2$.
A similar result holds for the region near the boundary of phase space
at $E_e = m_b/2$.  The differential cross section in eq.~\Vviii\ must be
smeared over a region of electron energies $\Delta E_e >
\lqcd$ to be physically meaningful.
Away from the boundaries $E_e = m_b/2$ and $q^2 = 2 m_b E_e$ of phase
space some smearing may also be required.  As one varies $E_e$ (at fixed
$q^2$)  new thresholds are encountered corresponding to different $M_X$
in eq.~\IXi.  However, here the situation is likely to be similar to
$R(s) = \sigma (e^+ e^- \rightarrow$  hadrons)/$\sigma (e^+ e^-
\rightarrow \mu^+ \mu^-$) where at large $s$ the density of hadronic
states is so large that the size of the region $s$  that must be
averaged over becomes negligibly small.

In semileptonic $B$ decay the endpoint region of electron energies,
where $B\rightarrow X_c e\bar\nu_e$ decays are forbidden, is of great
interest because of its potential use to determine $|V_{ub}|$.  However,
in this endpoint region $2.2\, {\rm GeV} < E_e < 2.6\, {\rm GeV}$ our
predictions are not very
useful.  Very near $q^2 = M_B^2$ our results don't apply and away from this
value of $q^2$  the averaging of electron energies over a region $\Delta
E_e > \lqcd$ makes it impossible to isolate the endpoint part of the
electron spectrum. This is discussed more quantitatively in the next
section.

\newsec{Numerical Results}

The $1/m_b^2$ corrections to the decay distribution for semileptonic $H_b$
decay have been computed in terms of the parameters $G_b$ and $K_b$. These
parameters depend on the hadron $H_b$, and will be denoted by $G_b(H_b)$ and
$K_b(H_b)$.
The parameter $G_b$ is the leading operator that breaks the heavy quark spin
symmetry, so its matrix element can be determined in terms of the hyperfine
spin-splittings within heavy quark multiplets. For example, $G_b(\Lambda_b)=0$
and
\eqn\Xi{
m_b G_b(B)= -\bfrac34\left[M(B^*)-M(B)\right],\qquad
m_b G_b(B^*)= \bfrac14\left[M(B^*)-M(B)\right].
}
The experimental value of 46~MeV for the $B^*-B$ mass difference determines
$G_b(B)$ to be $-0.0065$ (where $m_b$ can be equated with the hadron
mass to this
order). The parameter $K_b(B)$ cannot be simply determined. Quark model
estimates suggest that $K_b(B)$ is approximately 0.01. The size of the
$1/m_b^2$ correction to the lepton spectrum $d\Gamma/dy$ for
$B\rightarrow X_c e \bar \nu_e$
decay is plotted in fig.~5 for $G_b(B)=-0.0065$ and $K_b(B)=0.01$.
The plot shows the ratio of the distribution eq.~\Vvi\ to the free quark
decay spectrum
without any smearing. Fig.~6 shows the same result as a percentage
correction, so that one can see how the corrections become large only
near the endpoint. The results must be
averaged in $y$ over a region sufficiently large that the correction terms are
small.  The $1/m_c^2$ corrections to
charm decay are a factor of $(m_b/m_c)^2$ larger than the corrections for
b-decay.

\insertfig{\centerline{Figure 5}\medskip The ratio of the lepton
spectrum for $B\rightarrow X_c e\bar \nu_e$ decay including
$1/m_b^2$ corrections with $G_b=-0.0065$ and $K_b=0.01$ to the free
quark $b \rightarrow c$ decay spectrum. Note the logarithmic scale.}
{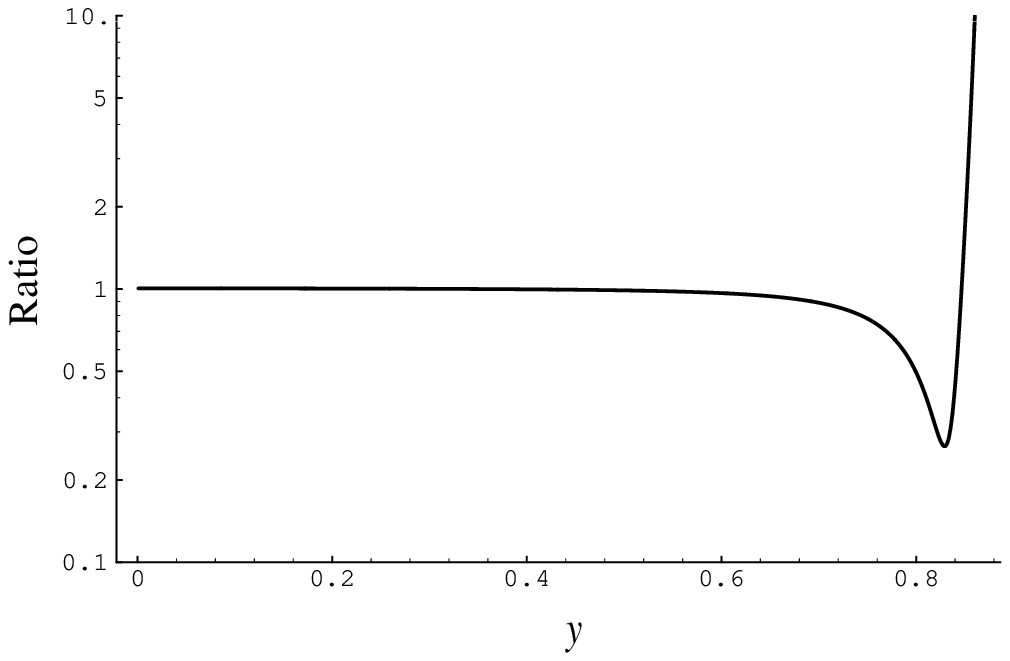}

\insertfig{\centerline{Figure 6}\medskip
\centerline{The same spectrum as fig.~5,
plotted as a percentage correction on a linear scale.}}{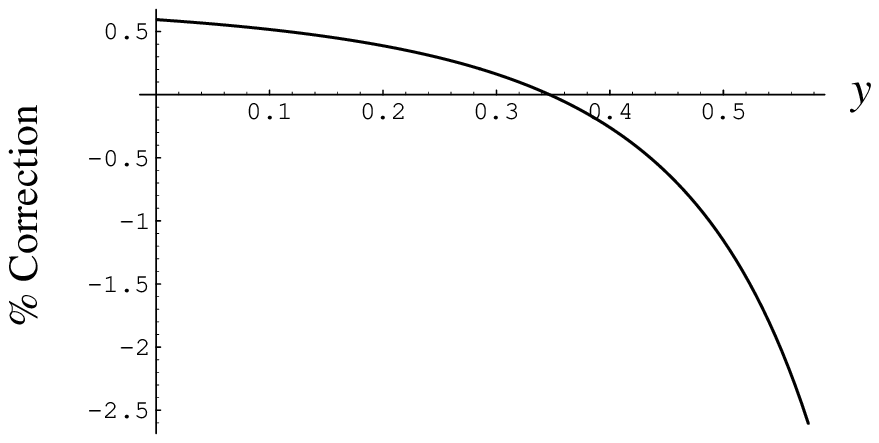}

To better understand the size of the $1/m_b^2$ corrections, it
is interesting
to plot the $B\rightarrow X_c e \bar\nu_e$ lepton spectrum including $1/m_b^2$
corrections without smearing. This
is illustrated in fig.~7, where the free quark $b\rightarrow c$ decay
and $b\rightarrow u$ decay
electron spectra are compared to the $B\rightarrow X_c e \bar\nu_e$
decay spectrum
including $1/m_b^2$ corrections. The  peak near the endpoint
for $B\rightarrow X_c e \bar \nu_e$ decay gives an indication of
the minimum size of the smearing region that must be used before the QCD
calculation is valid. The peak must be smeared over a large enough
region that it produces a small correction to the decay spectrum. This
indicates that smearing region should be at least $\Delta y=0.2$, which
corresponds to a lepton energy of around 500~MeV.

\insertfig{\centerline{Figure 7}\medskip The electron spectrum for free quark
$b\rightarrow c$ decay (dashed line), free quark $b\rightarrow u$ decay
(grey line), and  $B\rightarrow X_c e \bar\nu_e$ decay including
$1/m_b^2$ corrections
(solid line) with $G_b=-0.0065$ and $K_b=0.01$. The QCD calculation of the
$1/m_b^2$ corrections has not
been smeared.}{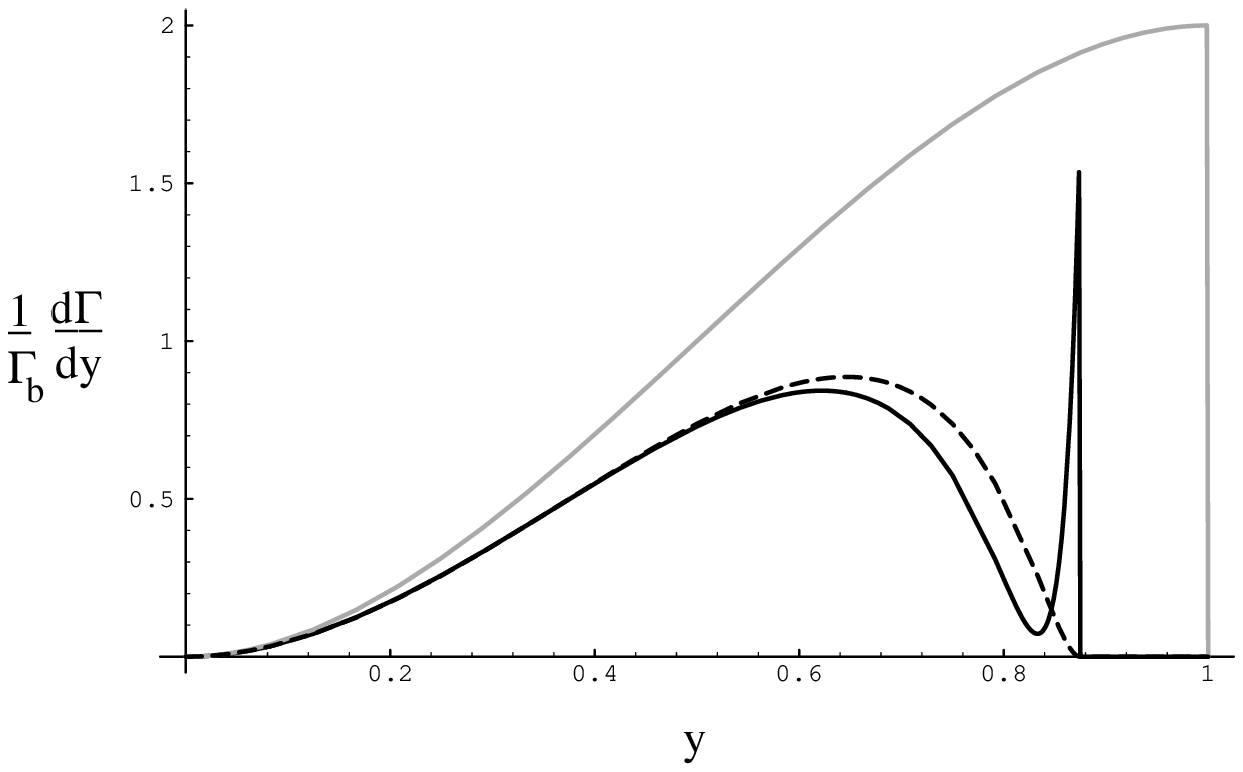}

The dominant uncertainty in the extraction of
$V_{ub}$ is the shape of the endpoint of the $B\rightarrow X_u
e\bar\nu_e$ lepton
spectrum. It is difficult to estimate the size of the smearing necessary
in QCD from the unsmeared spectrum eq.~\Vx\ since it contains
$\delta$-function singularities at the endpoint. Figure~8 shows the
smeared lepton decay distribution using Gaussian smearing in $y$ with
different widths. The smearing extends the spectrum beyond the parton
model endpoint $y=1$, but the curves are only plotted for $0\le y\le1$.
For simplicity, only the $\delta$-functions in eq.~\Vx\ have been
smeared, and then added to the remaining terms.
Clearly, the dip in the spectrum for a smearing width of
$\Delta y=0.1$ is unphysical. The curves in fig.~8 indicate that the
minimum smearing width in $\Delta y$ is around 0.2, which corresponds
to a lepton energy width of 500~MeV.

\insertfig{\centerline{Figure 8}\medskip Plot of the smeared corrections for
$B\rightarrow X_u e \bar\nu_e$ decay for $G_b=-0.0065$ and $K_b=0.01$.
The plots are
shown only in the region $0\le y \le 1$. The curves are the free quark
spectrum (solid grey), the $1/m_b^2$ corrected spectrum smeared
over $\Delta y=0.1$ (dashed grey), 0.2 (solid black) and 0.5 (dashed
black).}{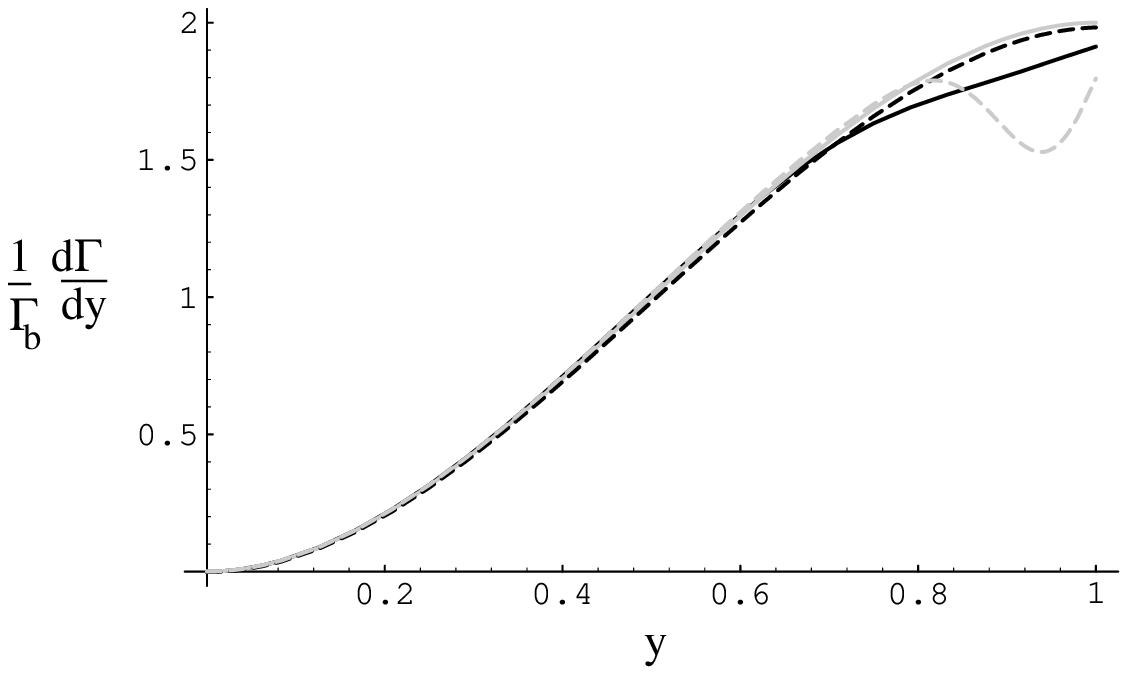}

Even though the parameters $K_b(H_b)$ are not known except by model
calculations, one can determine the difference of $K_b$ between two hadrons.
For example, the masses of the $\Lambda_b$ baryon and the $B$ and $B^*$ mesons
are
\eqn\Xii{\eqalign{
M(\Lambda_b) &= m_b + \bar\Lambda(\Lambda_b) + m_b K_b(\Lambda_b),\cr
M(B) &= m_b + \bar\Lambda(B) + m_b K_b(B) + m_b G_b(B),\cr
M(B^*) &= m_b + \bar\Lambda(B) + m_b K_b(B) - \bfrac13 m_b G_b(B),\cr
}}
including all corrections to order $1/m_b$ in the mass. We have used the heavy
quark spin symmetry relations $G_b(B)=-3G_b(B^*)$ and $K_b(B)=K_b(B^*)$. The
difference between the $\Lambda_b$ mass and the average meson mass $M(B)_{\rm
avg} = (M(B)+3M(B^*))/4$ is
\eqn\Xiii{
M(\Lambda_b)-M(B)_{\rm
avg}=\bar\Lambda(\Lambda_b)-\bar\Lambda(B)+m_b\left(K_b(\Lambda_b)-
K_b(B)\right).
}
A similar expression holds for the $D$ mesons and the $\Lambda_c$ baryon. Heavy
quark symmetry implies that $\bar\Lambda(H_c)=\bar\Lambda(H_b)$, and $m_b^2
K_b(H_b)=m_c^2 K_c(H_c)$, so that
\eqn\Xiv{
\left[M(\Lambda_c)-M(D)_{\rm avg}\right]-\left[M(\Lambda_b)-M(B)_{\rm
avg}\right] =m_b \left({m_b\over m_c}-1\right)\left(K_b(\Lambda_b)-
K_b(B)\right).
}
Equating the quark masses with the meson masses to this order gives
$K_b(\Lambda_b)- K_b(B)=-0.002\pm0.006$, using the present value of
$5641\pm50$~MeV for the $\Lambda_b$ mass.

The differences between $G_b$ and $K_b$ for different hadrons gives
predictions for the differences in the semileptonic decay distributions.
The order $\alpha_s$ corrections due to gluon radiation that have been
computed~\cabibbo\ali\accmm\ are the same for all hadrons, since they
correct the free quark decay formula, and cancel in the difference. Thus
the difference is known to corrections of order $\alpha_s(m_b)/m_b$ and
$1/m_b^3$. For example, one can compute the differences in the total
semileptonic decay widths for $H_b\rightarrow X_u e \bar\nu_e$ decay,
\eqn\Xv{
{\Gamma(H_b)-\Gamma(H^\prime_b)\over \Gamma(H_b)+\Gamma(H^\prime_b)} =
\bfrac32 \left(G_b(H_b)-G_b(H^\prime_b)\right)
-\bfrac12\left(K_b(H_b)-K_b(H^\prime_b)\right).
}
This gives
\eqn\Xvi{
{\Gamma(\Lambda_b)-\Gamma(B)\over \Gamma(B)} =
\bfrac94{\left(M(B^*)-M(B)\right)\over
M(B)}-\left(K_b(\Lambda_b)-K_b(B)\right)=0.018\pm 0.006.
}
Similarly, the decay width differences for $H_c\rightarrow X_d e^+
\nu_e$ decay are
\eqn\Xvii{
{\Gamma(\Lambda_c)-\Gamma(D)\over \Gamma(D)} =
\bfrac94{\left(M(D^*)-M(D)\right)\over
M(D)}-\left(K_c(\Lambda_c)-K_c(D)\right)=0.16\pm0.04.
}
The uncertainties in eqs.~\Xvi\ and \Xvii\ are due to the $\pm50$~MeV
uncertainty in the $\Lambda_b$ mass. The differences of the decay widths
for the decay modes $H_b\rightarrow X_c e\bar\nu_e$
and $H_c\rightarrow X_s e^+\nu_e$ can be computed similarly using
eqs.~\Vxi\ and \VIIiv.

\newsec{Conclusions}

One of the important points of the analysis in this paper is that the decay
distribution is determined by the quark mass $m_b$, rather than the
hadron mass $M_{H_b}$. Thus the free quark decay rate $\Gamma_b$ depends on
$m_b^5$, not $M_{H_b}^5$. The difference between the two masses is the
$\bar\Lambda$
parameter of the heavy quark theory. There should have been $1/m_b$ corrections
proportional to $\bar\Lambda$ if a free quark decay model
with the decay rate given by $M_{H_b}$ was appropriate. Corrections of
this form are absent in QCD.

The endpoint region of the inclusive lepton spectrum in semileptonic
b-decay is important for the extraction of $V_{ub}$. The QCD calculation
near the endpoint must be smeared over a large region (around 500~MeV)
before it can be
compared to experiment. This region is larger than
the difference in the endpoints of the $b\rightarrow u$ and
$b\rightarrow c$ decays. This means that the extraction of $V_{ub}$ still
requires modeling the endpoint. The QCD computation does provide some
constraints on the model. Any model which has $1/m_b$ corrections, or
has a decay distribution given by hadron kinematics instead of
quark kinematics, is in contradiction with QCD.

We have presented results for decays of hadrons containing a b or c
quark. Formally, perturbative corrections are of the form
$\alpha_s(m_b)$ or $\alpha_s(m_c)$. However, since the final state in
free quark decay is three-body, perturbative corrections to the total
semileptonic decay rate may be better represented by $\alpha_s(m_b/3)$ or
$\alpha_s(m_c/3)$. Only a higher order perturbative calculation can
resolve this issue. Because of this, we don't have confidence in
applying the results of sec.~7\ to the decays of hadrons containing a
c-quark.

\bigskip\centerline{{\bf Acknowledgements}}

While this paper was being written, we received a preprint by
Blok, Koyrakh, Shifman and Vainshtein~\ref\blok{B.~Blok,
L.~Koyrakh, M.~Shifman and A.I.~Vainshtein, ITP preprint NSF-ITP-93-68
{\tt [hep-ph/9307247]}} that also
computes the double differential decay distributions for unpolarized
b-hadrons. We thank B.~Blok for giving us a copy of this paper prior to
submission, and for discussing their results. T.~Mannel is also
preparing a manuscript on the double differential decay distributions
for $B$-decay~\ref\mannel{T. Mannel, IKDA-93/26 {\tt [hep-ph/9308262]}}.
We thank him for
discussing his work prior to publication.
We would also like to thank A.~Falk, M.E.~Luke and M.~Savage for
discussions. A.M. would like to thank the Aspen Center for Physics for
hospitality while part of
this work was completed. This work was supported in part by the Department of
Energy under grants DOE-FG03-90ER40546 and DEAC-03-81ER40050, by a TNLRC
grant RGFY93-206, and by a
Presidential Young Investigator award PHY-8958081 from the National
Science Foundation.

\listrefs
\end